\documentclass{CUP-JNL-DTM}%

\usepackage{graphicx}
\usepackage{multicol,multirow}
\usepackage{amsmath,amssymb,amsfonts}
\usepackage{mathrsfs}
\usepackage{amsthm}
\usepackage{rotating}
\usepackage{appendix}
\usepackage[numbers]{natbib}
\usepackage{ifpdf}
\usepackage[T1]{fontenc}
\usepackage{newtxtext} 
\usepackage{newtxmath}
\usepackage{textcomp}
\usepackage{xcolor}
\usepackage{lipsum}
\usepackage[noend]{algpseudocode}

\usepackage{algorithm}
\usepackage{verbatim}
\usepackage[colorlinks,allcolors=blue]{hyperref}

\newtheorem{theorem}{Theorem}[section]
\newtheorem{lemma}[theorem]{Lemma}
\newtheorem{definition}[theorem]{Definition}

\theoremstyle{definition}

\numberwithin{equation}{section}

\begin{document}
	
	\begin{Frontmatter}
		
		\title[Article Title]{Multivariate Intrinsic Local Polynomial Regression on Isometric Riemannian Manifolds: Applications to Positive Definite Data.  }
		
		\author[1,2]{Ronaldo Garc\'ia Reyes}\orcid{0000-0001-6206-5852}
		\author[2]{Ying Wang}\orcid{0000-0002-1947-3313}
		\author[2]{Min Li}\orcid{0000-0002-0835-2129}
		\author[1]{Marlis Ontivero-Ortega's }\orcid{0000-0003-0084-8274}
		\author[1,2]{Deirel Paz-Linares}\orcid{0000-0002-2144-8184}
		\author[1]{L\'idice Gal\'an Garc\'ia}\orcid{0000-0001-6400-6442}
		\author[1,2]{Pedro Antonio Valdes Sosa}\orcid{0000-0001-5485-2661}
		
		\authormark{Author Name1 \textit{et al}.}
		
		\address[1]{\orgdiv{Clinical Hospital of Chengdu Brain Science Institute}, \orgname{University of Electronic Science and Technology of China}, \orgaddress{\city{Chengdu}, \postcode{610054}, \state{China}}. \email{ronaldo@neuroinformatics-collaboratory.org,pedro.valdes@neuroinformatics-collaboratory.org}}
		
		\address[2]{\orgdiv{Neuroinformatics}, \orgname{Cuban Neurosciences Center}, \orgaddress{\city{Havana}, \postcode{11300}, \state{Havana},  \country{Cuba}}.}

		
		\authormark{ Garc\'ia  et al., 2023}
		
		
		
		\abstract{\small 	The paper introduces a novel non-parametric Riemannian regression method using Isometric Riemannian Manifolds (IRMs). The proposed technique, Intrinsic Local Polynomial Regression on IRMs (ILPR-IRMs), enables global data mapping between Riemannian manifolds while preserving underlying geometries. The ILPR method is generalized to handle multivariate covariates on any Riemannian manifold and isometry. Specifically, for manifolds equipped with Euclidean Pullback Metrics (EPMs), a closed analytical formula is derived for the multivariate ILPR (ILPR-EPM). Asymptotic statistical properties of the ILPR-EPM for the multivariate local linear case are established, including a formula for the asymptotic bias, establishing estimator consistency. The paper showcases possible applications of the method by focusing on a group of Riemannian metrics on the Symmetric Positive Definite (SPD) manifold, which arises in machine learning and neuroscience. It is demonstrated that several metrics on the SPD manifold are EPMs, resulting in a closed analytical expression for the multivariate ILPR estimator on the SPD manifold. The paper evaluates the ILPR estimator's performance under two specific EPMs, Log-Cholesky and Log-Euclidean, on simulated data on the SPD manifold and compares it with extrinsic LPR using the Affine-Invariant when scaling the manifold and covariate dimension. The results show that the ILPR using the Log-Cholesky metric is computationally faster and provides a better trade-off between error and time than other metrics. Finally, the Log-Cholesky metric on the SPD manifold is employed to implement an efficient and intrinsic version of Rie-SNE for visualizing high-dimensional SPD data. The code for implementing ILPR-EPMs and other relevant calculations is available on the GitHub page URL: \url{https://github.com/ronald1129/Multivariate-Intrinsic-Local-Polynomial-Regression-on-Isometric-Riemannian-Manifolds.git}.}  
		
	\end{Frontmatter}
	
	{
		\small\localtableofcontents}
	
	\section{Introduction}
	\small

	Regression models the relationship between a dependent variable  ${Y}_i \in  \mathcal{Y}$ and an independent variable $X_i \in \mathcal{X}$. This relationship is inferred from a dataset $\mathbf{\Sigma} =\{(X_i,Y_i)\in \mathcal{X}\times \mathcal{Y}: i=1,2,\cdots N\}$. The goal is to find the best-fitting curve or hyper-surface that describes the relationship between the variables.
	Non-parametric regression makes minimal assumptions about the functional form of the relationship between the variables. In Euclidean space,  $\mathcal{X}\times  \mathcal{Y} = \mathbb{R}^p\times \mathbb{R}^n$, if $\mathbf{\Sigma}$ correspond to a set of independent and identically realization of a sample population $({X},{Y})$,  non-parametric regression aims to determine a smooth regression function $m(\cdot)$ that satisfies 
	\begin{align}\label{regression_statment}
		m({x}) = \mathbb{E}[Y|{X} = {x}],
	\end{align}
	without parametrically  specifying the shape of $m(\cdot)$. Unlike parametric regression, which assumes a specific mathematical equation (e.g.,  $m({X})= {\alpha}_0+{\alpha}_1^T{X}$), non-parametric regression allows for a more flexible approach where the shape of $m(\cdot)$ is inferred from the data $\mathbf{\Sigma}$ \cite{ruppert1994multivariate,fan2018local,wand1994kernel}. 
	
	One specific type of non-parametric regression in the Euclidean space is Local Polynomial Regression ({LPR}). This method fits a Local Polynomial (LP) regression model to the data in a local neighborhood around each data point instead of using a single global polynomial to fit the entire dataset. LPR allows a more adaptive model to handle complex nonlinear relationships in the data. In precise terms, LPR of order $k$ is a non-parametric method that consists of estimates a $k+1$ times differentiable regression function $m(x)$  and its derivatives  $\partial^j_{x}m(X), \,\, 0<j\leq k$. In contrast to global polynomial regression, LPR effectively handles cases where the relationship between the variables is heterogeneous across the dataset. Other non-parametric regressions in Euclidean space are Kernel regression and Spline Regression  \cite{ruppert1994multivariate,fan2018local,wand1994kernel,avery2013literature}. 
	
	Riemannian value data, $ \mathcal{Y}=\mathcal{M}_n\ne \mathbb{R}^n$,  arise
	in fields such as Medical Imagen, Computational Anatomy, Machine Learning, Computer Vision, Finance, and others \cite{lepore2007generalized,li2022harmonized,barachant2013classification,tuzel2008pedestrian}. In contrast to Euclidean space, a Riemannian manifold $(\mathcal{M}_n, g)$ is a nonlinear mathematical structure that exhibits local similarities with Euclidean space. Specifically, at each point, it can be approximated by a tangent space, essentially a Euclidean space \cite{lee2018introduction,you2021re}. The nonlinearity implies that the structure is not flat, and the linear combinations of elements in $\mathcal{M}_n$ are not granted to remain in the manifold. Thus, a simple estimator of central tendency for data on a Riemannian Manifold $(\mathcal{M}_n, g)$ like the Euclidean Mean ($\widehat{Y} = \sum Y_i /N$)  is generally not consistent with the geometrical structure of the data ($\widehat{Y}\notin \mathcal{M}_n$). The appropriate manner of computing the mean of Riemannian value data necessitates the utilization of the Karcher Mean \cite{pennec2006riemannian}. 
	The same situation occurs when performing regression on Riemannian value data. The relationships between data points in a manifold-valued dataset $\mathcal{X}\times \mathcal{Y} = \mathbb{R}^p\times\mathcal{M}_n$ may be complex and nonlinear and depend on the geometry of the underlying manifold, making traditional Euclidean-based regression techniques as the LPR unsuitable. In such cases, Riemannian manifold regression methods are required to model the relationships in the data accurately \cite{yuan2012local,petersen2019frechet,dryden2009non}.
	
	The main feature of the Riemannian regression approach is to rewrite the classical regression techniques in terms of Riemannian operations like the exponential map, logarithm map, parallel transport, and intrinsic distance (these concepts are defined in the section \ref{basic}). This approach allows us to avoid the regression estimators' inconsistency with the data's geometrical structure \cite{pennec2006riemannian}. In the context of regression on a Riemannian manifold, there are two primary approaches: extrinsic and intrinsic. An intrinsic method refers to one where the unknown regression function $m(X)$ belongs to the same manifold or an equivalent Riemannian manifold that preserves all of the original geometrical information, i.e., $m(X)\in\mathcal{M}$. On the other hand, for the extrinsic approach, $m(X)$ belongs to a higher dimensional Euclidean space that does not conserve the underline geometric of $\mathcal{M}$.
	

	The extrinsic approach comprises two steps: firstly, mapping the Riemannian manifold value data into a higher-dimensional Euclidean space, and secondly, performing regression on the mapped data using classical regression methods. While this approach is straightforward, the mapped data may not precisely reflect the genuine relationships present in the data \cite{li2022harmonized,lin2017extrinsic,dai2018principal,lin2019intrinsic,lee2021robust,lin2019extrinsic,you2021re}.
	
	Infinite ways exist to equip a Manifold $\mathcal{M}$ with a metric $g$. For each of these metrics, there are infinite ways of selecting embedding maps into a higher-dimensional Euclidean space. Some authors, like Lin, use equivariant embeddings to respect the underlying geometry of the Riemannian Manifold as much as possible \cite{lin2017extrinsic}. In practice, many authors in the literature define these embeddings by using the logarithm map at a convenient point in the manifold \cite{li2022harmonized,lin2017extrinsic}. A point is convenient if it contains as much information as possible about the variability of the data. A typical selection for this point is the Karcher Mean of the data. 
	
	Once the data is mapped into the high-dimensional space, we can choose any known classical regression techniques to perform regression. For example, we can apply Linear Regression, Kernel regression, Local Polynomial Regression, Spline Regression, or Additive Regression for handling multiple descriptors. The choice of the regression method on the High Dimensional Euclidean space depends on the problem we want to solve. Finally, we map the regression result into the original Riemannian Manifold using the "local inverse" of the embedding function \cite{lin2017extrinsic,you2021re}. In the context of the Symmetric Positive Definite (SPD) Manifold, the Affine-Invariant metric is the preferred choice due to its congruence invariance property, as evidenced by \cite{barachant2010riemannian,barachant2013classification,li2022harmonized,sabbagh2020predictive,arsigny2006log}. A widely used approach in the literature is to apply Extrinsic Local Polynomial Regression under the Affine-Invariant metric which allows for non-parametric local polynomial regression \cite{lin2017extrinsic,li2022harmonized,you2021re}.
	
	The intrinsic approach to regression on a Riemannian manifold involves using Riemannian-based methods to perform regression directly. This approach accurately captures complex relationships in the data. Among the intrinsic methods, Multivariate General Linear Models (MGLM) extend the classical parametric linear regression method in Euclidean space to Riemannian manifolds \cite{kim2014multivariate}. Additive Regression Models (ARM) on the Tangent Space extend the classical results in Additive Regression to a Riemannian Manifold with a Lie structure \cite{fan2018local,lin2022additive}. Fr\'echet Regression (FR) extends non-parametric regression methods from Euclidean Space to Metric Spaces. This type of regression employs only the distance between the elements and is valid on general Metric Spaces where Taylor expansions may not be possible \cite{petersen2019frechet}. Finally, Intrinsic Local Polynomial Regression (ILPR) is a non-parametric regression method that extends the classical LPR on Euclidean space to Riemannian Manifolds \cite{yuan2012local}.	The main difficulty posed by the intrinsic approach lies in the complexity of the resulting regression optimization problems. Achieving an exact solution is frequently impractical, and numerical techniques are often employed, which can be computationally costly and reduce the precision of the intrinsic methods \cite{petersen2019frechet,yuan2012local}. Consequently, novel theoretical instruments are essential to simplify the mathematical intricacies and facilitate a more direct resolution of intrinsic optimization problems.
	
	This paper focuses on developing new theoretical results for the multivariate Intrinsic Local Polynomial Regression (ILPR) introduced by Yuan and colleagues \cite{yuan2012local}. The current state-of-the-art in ILPR lacks a formal definition capable of handling multivariate covariates on an arbitrary Riemannian manifold and a general expression for computing ILPR for new metrics on the Symmetric Positive Definite (SPD) manifold, such as the Log-Cholesky metric \cite{lin2019riemannian,dryden2009non}. Additionally, there is a lack of closed analytic expression for ILPR and general results regarding its asymptotic properties on arbitrary Riemannian manifolds.
	
	We propose a solution to these issues by formally defining multivariate ILPR on arbitrary Riemannian manifolds and studying the effects of isometries on the estimation. Riemannian isometries are mathematical functions that preserve the geometric structure of a Riemannian manifold, allowing for the transformation of one Riemannian manifold into another while preserving distances and angles between points \cite{thanwerdas2021geometry}. Isometric Riemannian manifolds (IRMs) are a subset of Riemannian manifolds that share similar geometric properties through isometry. Thus, analyzing complicated Riemannian manifolds can be simplified by studying IRMs with more straightforward geometric structures.
	
	The main thrust is that intrinsic Riemannian regression methods, such as ILPR, can be simplified by expressing them on IRMs. This simplification allows for more straightforward solutions to regression problems on Riemannian Manifold. Therefore, this article aims to expand the analytical tools of ILPR beyond the SPD manifold and generalize it to multivariate cases on any Riemannian manifold. To achieve this, we introduce a new method for computing ILPR on IRMs (ILPR-IRMs), allowing global data mapping from one Riemannian manifold to another while preserving the underlying geometry. ILPR-IRMs enable the expression of regression in the most convenient Riemannian manifold. Once the solution to the reduced problem is found, IRM isometry allows us to recover the solution in the original Riemannian manifold. Thus, we address the following problems:
	
	\begin{enumerate}
		\item [(a)] We generalize the definition of multivariate ILPR to handle responses on any arbitrary Riemannian manifold that depend on multivariate covariates. This new definition establishes how an isometry on IRM affects ILPR in Theorem (\ref{MILPR}).
		
		\item [(b)] The exact analytical expression for multivariate ILPR on an arbitrary Riemannian manifold equipped with a Euclidean Pullback Metric (EPM) is derived in Theorem (\ref{teorema3}). The estimator has a closed expression reminiscent of LPR in Euclidean space.
		$\widehat{\alpha}_0(x;k,h) = f^{-1}\left( f(\mathbf{Y})\left((\mathbf{WX}^T(\mathbf{XWX}^T)^{-}e_0)\otimes I_n\right)\right)
		$.
		
		\item[(c)]The Log-Euclidean, Power-Euclidean, and Cholesky metrics are known to be EPMs. In Lemma (\ref{lemma1}), we demonstrate for the first time that the Log-Cholesky metric is also an EPM.	
		\item [(d)] In Theorem (\ref{teorema3}), we provide the exact analytical expression of ILPR on the SPD manifold equipped with an EPM. We compare this methodology using simulated data on the SPD manifold with Log-Cholesky and Log-Euclidean metrics and the established extrinsic approach using the Affine Invariant distance. These simulations are conducted in two different settings for data on SPD($3$) and SPD($n$) for varying dimensions of covariates. Based on the simulated data, performing ILPR using the Log-Cholesky metric is computationally faster and offers the best trade-off between Error-Time and other metrics considered. However, while the extrinsic Affine-Invariant metric is slightly more accurate, it requires significantly more computational time \ref{simulations}.   
		\item [(e)] We developed the Log-Cholesky Rie t-SNE, a dimensionality reduction technique based on the Rie-SNE method \cite{bergsson2022riesne}, to visualize the performance of ILPR on high-dimensional simulated regression data.
		\item[(f)] The code for implementing ILPR-EPMs, Log-Cholesky Rie t-SNE, and other relevant calculations is available on the GitHub page URL: \url{https://github.com/ronald1129/Multivariate-Intrinsic-Local-Polynomial-Regression-on-Isometric-Riemannian-Manifolds.git}.
	\end{enumerate}
	
	\newpage
	\section{\small General Notation}
	\begin{table}[h!]
		\small
		\TBL{\caption{ \small \bf Basic Notation }}
		{\begin{tabular*}{\textwidth}{@{\extracolsep{\fill}}llllll@{}}\toprule
				\label{Tab3}
				\TCH{\small \bf Manifold} & \TCH{\small \bf Description} & \TCH{\small \bf Bib}  \\\midrule
				\small	$\mathbb{R}^p$ &  \small Euclidean Space of dimension $p$ & --
				\\ \small	$\operatorname{M}_n(\mathbb{R})$ &  \small Space of ${n\times n}$ matrices with real components & --
				\\ \small	$\operatorname{GL}_n(\mathbb{R})$ &  \small $\operatorname{GL}_n(\mathbb{R})=\{Y\in \operatorname{M}_n(\mathbb{R}): \operatorname{det}(Y)\ne 0 \}$ & --
				\\ \small	$\mathcal{M}$ &  \small Arbitrary  manifold & \small  (\ref{basic})
				\\ \small	$\mathcal{T}_Y\mathcal{M}$ &  \small Tangent space of the manifold $ \mathcal{M}$ at $Y$, ($Y\in\mathcal{M}$)  & \small  (\ref{basic})
				\\ \small	$\mathcal{M}_n$ &  \small Submanifold of $\operatorname{M}_n(\mathbb{R})$ & \cite{kosinski2013differential}
				\\ \small	$\operatorname{Sym}_n$ &\small  $\operatorname{Sym}_n=\{Y\in \operatorname{M}_n(\mathbb{R}):Y=Y^T\}$    & \cite{arsigny2006log}
				\\ \small	$\operatorname{LT}_n$ & \small 	$\operatorname{LT}_n=\{Y\in \operatorname{M}_n(\mathbb{R}): \forall i<j, Y_{ij}=0 \}$   & \cite{lin2019riemannian}
				\\ \small	$\operatorname{Diag}_n$ & \small 	$\operatorname{Diag}_n=\{Y\in \operatorname{M}_n(\mathbb{R}): \forall i\ne j, Y_{ij}=0 \}$   & \cite{lin2019riemannian}
				\\ \small	$\operatorname{SPD}_n$ &  \small  	$\operatorname{SPD}_n =\{Y\in \operatorname{M}_n(\mathbb{R}): \forall {X}\in\mathbb{R}^n, {X}^TY{X}>0 \} $  & \cite{arsigny2006log}
				\\\botrule
				\toprule
				\TCH{\small \bf Matrix Operation} & \TCH{\small $Y,Y_1,Y_2\in \operatorname{M}_n(\mathbb{R})$ \& $x,X\in\mathbb{R}^p$} & \TCH{\small --}  \\\midrule 
				\small	$O_n$, $I_n$ &  \small  Zero and Identity matrix of $\text{M}_n(\mathbb{R})$  & --
				\\ \small	$Y^T$, $Y^{-1}$  &  \small Transpose matrix and  Inverse matrix of  non singular  matrix $Y$  & --
				\\ \small	$\operatorname{col}_i(Y)$, $\operatorname{row}_i(Y)$ &  \small  Column $i$ of   $Y$,  Row $i$ of   $Y$  & --	
				\\ \small	$\operatorname{colb}_i(Y)$, $\operatorname{rowb}_i(Y)$ &  \small  Column or Row of Blocks $i$ of the block matrix  $Y$ & \cite{koning1991block}	
				\\ \small	$\operatorname{tr}(Y)$ &  \small Trace operator, $\operatorname{tr}(Y_{n\times n})=\sum_{1\leq i\leq n} Y_{ii}$  & --			
				\\ \small	$\operatorname{exp}(Y)$ &  \small Matrix exponential, $\operatorname{exp}(Y)=\sum_{i=0}^{\infty} Y^n/n!$  & \cite{arsigny2006log}
				\\ \small	$\operatorname{log}(Y)$ &  \small Matrix Logarithm, $\operatorname{log} (Y) =\operatorname{exp}^{-1}(Y)$  & \cite{arsigny2006log}
				\\ \small	$Y_1\otimes Y_2$ &  \small Kronecker product of $Y_1$ and $Y_2$   & \cite{van2000ubiquitous}
				\\ \small	$Y_1\boxtimes Y_2$ &  \small Tracy-Singh Block Kronecker product of $Y_1$ and $Y_2$ without partitioning $Y_1$   & \cite{koning1991block}
				\\ \small	$Y_1:Y_2$ &  \small Frobenius product,  $Y_1:Y_2=\operatorname{tr}(Y_1Y^T_2)$   & \cite{horn2012matrix}
				\\ \small	$||Y||_2$ &  \small Frobenius norm,  $||Y||_2=\sqrt{Y:Y}$   & \cite{horn2012matrix}
				\\ \small	$Y_{n\times m}^{\otimes j}$ &  \small Kronecker power,  $Y_{n\times m}^{\otimes j}=Y_{n\times m}\otimes Y_{n\times m}^{\otimes (j-1)}$, $Y_{n\times m}^{\otimes 0} =I_{m}$   & \cite{van2010consistent}
				\\ \small	$\partial^j_{x}Y(X) $ &  \small Partial derivative of a vector value matrix function of order $j$   & \cite{van2010consistent}
				\\ \small	$\mathcal{L}(Y)$ &  \small Cholesky decomposition of a SPD matrix $Y$  & \cite{lin2019riemannian}
				\\ \small	$\mathbb{D}(Y)$, $\lfloor Y\rfloor$ &  \small Diagonal part of $Y$, Strictly lower triangular part of $Y$    &  \cite{lin2019riemannian}
				\\ \small	$\operatorname{pow}_\tau(Y)$ &  \small $\operatorname{pow}_{\tau}(Y)=\exp(\tau\log Y)$    &  \cite{dryden2009non}
				\\ \small	$(Y)_{1/2}$ &  \small $(Y)_{1/2} = \lfloor Y \rfloor + \mathbb{D}(Y)/2$    &  \cite{lin2019riemannian}
				\\\botrule
				\toprule
				\TCH{\small \bf Abbrebiatons} & \TCH{\small --} & \TCH{\small --}  \\\midrule
				\small IRM &  
				\small Isometric Riemannian Manifold  & \small   (\ref{DeformedRiemannianManifold})
				\\\small	EPM &  
				\small  Euclidean Pullback Metric  & \small   (\ref{def2})
				\\\small	LPR &  
				\small Multivariate Local Polynomial Regression in Euclidean Space  & \small   (\ref{lpr})
				\\\small	ILPR &  
				\small Multivariate Intrinsic Local Polynomial Regression    & \small   (\ref{genemilpr})
				\\\small	ILE &  
				\small   Intrinsic Local Estimator & \small   (\ref{def3})
				\\\botrule
			\end{tabular*}}
		\end{table}
		\newpage
		\newpage
		
		\section{Background}
	\subsection{Riemannian Manifolds} \label{basic}

	A manifold $\mathcal{M}$ is a topological space that locally resembles a Euclidean space. Specifically, for each element $Y \in \mathcal{M}$, it is possible to associate a Euclidean space called a tangent space, denoted as ${T}_Y\mathcal{M}$. Suppose we define in the tangent space a map $g_S: T_Y\mathcal{M}\times{T}_Y\mathcal{M}\rightarrow \mathbb{R}$ such that $g_Y$ is an inner product that varies smoothly concerning $Y\in \mathcal{M}$. In that case, the pair $(\mathcal{M}, g)$ constitutes a Riemannian manifold, with $g$ being a Riemannian metric \cite{lee2018introduction,you2021re}.
	
	On a manifold, a curve is a differentiable map $\gamma:[t_0,t_1]\rightarrow \mathcal{M}$ such that for all $t\in [t_0,t_1]$, $\gamma'(t) \in T_{\gamma(t)}\mathcal{M}$. The geometric structure defined on a Riemannian manifold $(\mathcal{M}, g)$, along with the concept of a curve, allows us to define the geodesic distance between two elements $Y_1$ and $Y_2$ in $\mathcal{M}$, denoted by $\operatorname{dist}_g(Y_1, Y_2)$, as the length of the shortest path between $Y_1$ and $Y_2$
	\begin{eqnarray}\label{length}
		\begin{split}
			&\operatorname{dist}_g(Y_1,Y_2)=\min_{\gamma\in\tilde{\gamma}_{t_0,t_1}(Y_1,Y_2)}\int_{t_0}^{t_1}dt\sqrt{g_{\gamma(t)}(\gamma'(t), \gamma'(t))},
		\end{split}
	\end{eqnarray}
	where $\tilde{\gamma}_{t_0,t_1}(Y_1,Y_2)$ is the set of all curves such that $\gamma(t_0)=S$ and $\gamma(t_1)=K$.  A geodesic is a smooth curve on $\mathcal{M}$ whose tangent vector $\gamma'$ does not change length or direction as one moves along the curve. For $V\in T_Y\mathcal{M}$ there is a unique geodesic, denoted by $\gamma_{Y,V}(t)$, whose domain contains $[0,1]$ such that $\gamma_{Y,V}(0)=S$  and $\gamma'_{Y,V}(0)=V$.
	
	In general, a Riemannian Manifold is a nonlinear space, and unlike a Euclidean space, linear combinations may not remain in the Manifold. However, three essential operations can be defined to manipulate and study the Manifolds: the Exponential map, the Logarithmic map, and the Parallel Transport. The Exponential map, denoted by $\operatorname{exp}_{Y}:\mathcal{T}_Y\mathcal{M}\rightarrow \mathcal{M}$, is defined as $\operatorname{exp}_{Y}(V)=\gamma_{Y, V}(1)$, where $\gamma_{Y,V}$ is the unique geodesic with initial condition $\gamma_{Y,V}(0) = Y$ and $\gamma'_{Y,V}(0) = V$. The Logarithmic Map, denoted by $\operatorname{log}_{Y}:\mathcal{M}\rightarrow T_Y\mathcal{M}$, is defined as the inverse of the Exponential Map, i.e., $\operatorname{log}_Y=\operatorname{exp}^{-1}_{Y}$. Finally, the Parallel Transport, denoted by $\pi_{Y_1\to Y_2}: T_{Y_1}\mathcal{M}\to T_{Y_2}\mathcal{M}$, is a linear map on a Riemannian manifold that moves a tangent vector $V\in T_{Y_1}\mathcal{M}$ along a curve without changing its direction or orientation. These operations are crucial for studying and manipulating Riemannian manifolds and have practical applications in fields such as machine learning, computer vision, and medical imaging $\text{\cite{lee2018introduction,you2021re,pennec2006riemannian,yuan2012local}}$.
	\subsection{Symmetric Positive Definite Manifold (SPD)}
	The Symmetric Positive Definite (SPD) Manifold is a mathematical object that arises in various applications, such as computer vision, signal processing, and machine learning. During this article, we will illustrate the analytical tools on this Manifold. It consists of a collection of symmetric, positive-definite matrices, which are commonly used to represent covariance or similarity information in data sets or, more precisely \cite{arsigny2006log,pennec2006riemannian,barachant2010riemannian,dryden2009non,lin2019riemannian}:
	\begin{align}
		\begin{split}
			\text{SPD}_n &= \{ Y: \text{M}_n(\mathbb{R}): \forall {X}\in \mathbb{R}^n: {X}^TY{X}>0\}.
			\\T_Y\operatorname{SPD}_n &= \{ Y: \text{M}_n(\mathbb{R}): Y=Y^T\}=\operatorname{Sym}_n.
		\end{split}
	\end{align} 
	When equipped with a Riemannian metric, the SPD Manifold becomes a Riemannian Manifold. Several Riemannian metrics can be considered on the SPD Manifold, and a subset of the most important one is summarized in Table  (\ref{Tab5}). 
	
	\begin{table*}[h!]
		\small
		\TBL{\caption{ \small \bf Riemannian Metrics on the SPD Manifold }}
		{\begin{tabular*}{\textwidth}{@{\extracolsep{\fill}}llllll@{}}\toprule
				\label{Tab5}
				\TCH{\small \bf Metric} & \TCH{\small {\bf Description} ($Y\in\text{SPD}_n$ \& $V\in T_Y\text{SPD}_n\simeq \operatorname{Sym}_n$)} & \TCH{\small \bf Bib}  \\\midrule
				\small	Euclidean &  \small   $g^E_Y(V,V)=||V||_2^2$   & --
				\\	\small	 Power-Euclidean &  \small   $g^{PE}_Y(V,V)=||{d}_Y\operatorname{pow}_\tau(V)||_2^2$   & \cite{dryden2009non}
				\\\small	Log-Euclidean &  
				\small   $g^{LE}_Y(V,V)=\||{d}_Y\log(V)||_2^2$   & \cite{arsigny2006log}
				\\\small Cholesky &  
				\small   $g^{C}_Y(V,V)=||{d}_Y\mathcal{L}(V)||_2^2$   & \cite{dryden2009non}
				\\\small	Log-Cholesky &  
				\small  $g^{LC}_Y(V,V) = ||\lfloor \mathcal{L}(Y)(\mathcal{L}(Y)^{-1}V\mathcal{L}(Y)^{-T})_{1/2}\rfloor||_2^2 +|| \mathbb{D}((\mathcal{L}(Y)^{-1}V\mathcal{L}(Y)^{-T})_{1/2})||_2^2$  & \cite{lin2019riemannian}
				\\\small	Affine-Invariant &  
				\small  $g^{AI}_Y(V,V) = \operatorname{tr}((Y^{-1}V)^2)$   & \cite{pennec2006riemannian}	
				\\\botrule
			\end{tabular*}}
		\end{table*}
		
		The SPD Manifold has different Riemannian metrics that can be used depending on the problem to solve. For example, when analyzing large datasets, Log-Cholesky and Log-Euclidean metrics are well-suited. However, sometimes in necessary to take into account congruence invariance property, then the Affine-Invariant metric is considered more suitable.

		\subsection{Isometric Riemannian Manifolds (IRM) and Euclidean Pulback Metrics (EPM)} \label{DeformedRiemannianManifold}
		\begin{figure}[h!]
			\small
			\centering
			\includegraphics[width=0.6\linewidth]{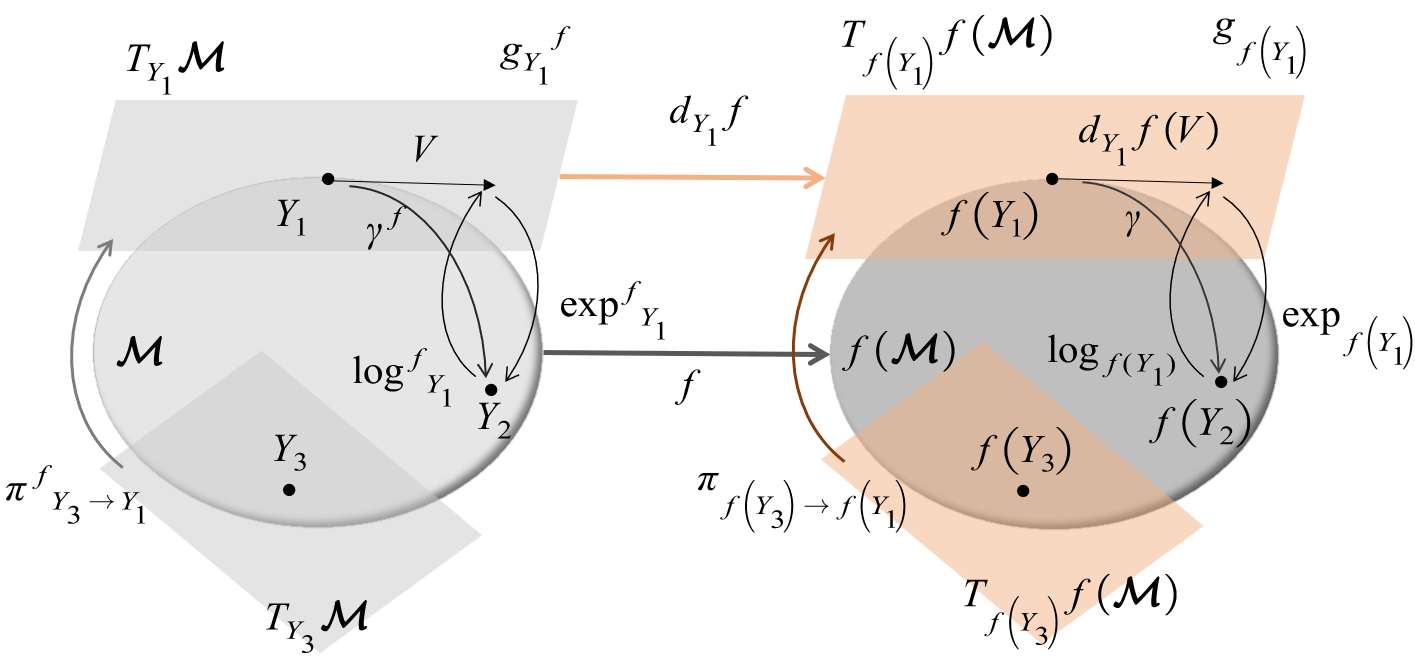}
			\caption{\small Visual representation of the relationships between Riemannian operations in a Manifold equipped with a Pullback Metric}
			\label{fig:relation}
		\end{figure}
		
		In various practical situations, it is possible to associate a manifold $\mathcal{M}_n$ with a simpler one $f(\mathcal{M}_n)$ through a diffeomorphism $f$. For instance, applying the logarithm to $\operatorname{SPD}_n$ yields the symmetric matrix manifold $\operatorname{Sym}_n$, which can be expressed as $\log(\operatorname{SPD}_n)=\operatorname{Sym}_n$ \cite{arsigny2006log}. Such situations are advantageous for extending a metric from one Manifold to another or simplifying a problem. Generally, given a diffeomorphism $f:\mathcal{M}_n\rightarrow f(\mathcal{M}_n)$, it is possible to define a new metric $g^f$ on the manifold $\mathcal{M}_n$ using $(f(\mathcal{M}_n),g)$ in a way that makes it a Riemannian manifold. This metric is known as the pullback metric \cite{lee2018introduction}:
		
		\begin{definition}[\bf Pullback Metric and Directional derivative]\label{pullb}
			Let $\mathcal{M}_n$ be a manifold and $(f(\mathcal{M}_n),g)$ be a Riemannian manifold. Consider a diffeomorphism $f:\mathcal{M}_n\rightarrow f(\mathcal{M}_n)$. The pullback metric of $g$ on $\mathcal{M}_n$ induced by $f$ is defined as follows. For $Y\in \mathcal{M}_n$ and $V\in T_Y\mathcal{M}_n$, we have
			\begin{align} \label{pullback}
				g^f_Y(V,V)=g{f(Y)}({d}_Yf(V),{d}_Yf(V)),
			\end{align}
			where ${d}_Yf(V):{T}_Y\mathcal{M}_n\to {T}_{f(Y)} f(\mathcal{M}n)$ is the directional derivative of $f$ at $Y$ in the direction $V$. That is,
			\begin{align}
				d_{Y}f(V) = \lim_{\epsilon\to 0} \epsilon^{-1}\left(f(Y+\epsilon V) - f(Y)\right).
			\end{align}
		\end{definition}
		This Definition allows the extension of all Riemannian operations from one Manifold to another, see Table  (\ref{Tab1}). To illustrate this concept, we can consider the SPD Manifold with the Log-Euclidean metric. Using Definition (\ref{pullb}), we can quickly check that the Log-Euclidean metric is the Pullback Metric of the Frobenius metric  \cite{arsigny2006log}.

		\begin{table*}[h!]
			\TBL{\caption{ \small \bf Relationships between Riemannian Operations in a Manifold equipped with a Pullback Metric \cite{thanwerdas2021geometry}}}
			{\begin{tabular*}{\textwidth}{@{\extracolsep{\fill}}llllll@{}}\toprule
					\label{Tab1}\small
					\TCH{\small \bf Riemannian Operation} & \TCH{ \small  $\mathcal{M}$} & \TCH{ \small $f(\mathcal{M})$} &\TCH{\small \bf Relationship ($Y,Y_1,Y_2\in\mathcal{M}$, $V\in T_Y\mathcal{M}$, $t\in\mathbb{R}$) } \\\midrule
					\small
					Metric & \small $g_Y^f$ & \small $g_{f(Y)}$ &\small $g^f_Y(V,V)=g_{f(S)}({d}_Yf(V),{d}_Yf(V))$\\
					Geodesic & \small $\gamma^f_{Y,V}$ & \small$\gamma_{f(Y),{d}_{Y}f(V)}$ &\small $\gamma^f_{Y,V}(t)=f^{-1}\left(\gamma_{f(Y),{d}_{Y}f(V)}(t)\right)$ \\ \small
					Exponential Map &\small $\operatorname{exp}^f_Y$ &\small $\exp_{f(Y)}$ &\small $\operatorname{exp}^f_{Y}(V)=f^{-1}\left(\exp_{f(Y)}\left({d}_{Y}f(V)\right)\right)$ \\\small
					Logarithm Map & \small$\operatorname{log}^f_Y$ &\small $\log_{f(Y)}$ & \small $\operatorname{log}^f_{Y_1}(Y_2)={d}_{f(Y_1)}f^{-1}\left(\log_{f(Y_1)}f(Y_2)\right)$
					\\\small Parallel Transport &\small$\pi^f_{Y_1\rightarrow Y_2}$ &\small $\pi_{f(Y_1)\rightarrow f(Y_2)}$ &\small   $\pi^f_{Y_1\rightarrow Y_2}V={d}_{f(Y_2)}f^{-1}\left(\pi_{f(Y_1)\rightarrow f(Y_2)}{d}_{Y_1}f(V)\right)$
					\\ \small Distance & \small$\operatorname{dist}_{g^f}$ &\small $\operatorname{dist}_{g}$ &\small $\operatorname{dist}_{g^f}(Y_1,Y_2)=\operatorname{dist}_g(f(Y_1),f(Y_2))$\\\botrule
				\end{tabular*}}
			\end{table*}
			
			In general, when two Riemannian Manifolds have equivalent geometric structures in the above sense, we can see each of these Riemannian Manifolds as the result of deforming the other thanks to a diffeomorphism $f$. This is what is known as Isometric Riemannian Manifold or, more precisely:
			
			\begin{definition}[\textbf{Isometric Riemannian Manifolds}]
				Let $(\mathcal{M}_n,g^f
				)$ and $(f(\mathcal{M}_n),g)$ be two Riemannian Manifolds. A diffeomorphism $f:\mathcal{M}_n\rightarrow f(\mathcal{M}_n)$ is an \textit{isometry} if it preserves the Riemannian metric, i.e., $g^f$ is the pullback metric of $g$ by $f$. Two Riemannian Manifolds $\mathcal{M}_n$ and $f(\mathcal{M}_n)$ are \textit{isometric} if there exists an isometry between them.
			\end{definition}
			
			One example of Isometric Riemannian Manifolds is the pair $(\operatorname{SPD}_n,g^{LE})$ and $(\operatorname{Sym}_n,g^{E})$. In this case, the function $\log$ is an isometry between them. This means that $\log(\operatorname{SPD}_n)$ is equal to $\operatorname{Sym}_n$, and the Riemannian metric $g^{LE}$ is obtained from $g^E$ by pulling it back using the $\log$ function \cite{arsigny2006log}. In other words, the two manifolds have the same underlying geometry and can be transformed into each other without changing the distances between points or angles between curves.
			
			One important concept discussed in this article is the Euclidean Pullback Metric (EPM). This type of Riemannian metric arises from the Euclidean metric's pullback via a diffeomorphism. In other words, given a diffeomorphism between two Riemannian manifolds, the EPM is obtained by "pulling back" the Euclidean metric from one Manifold to the other. This allows us to compare distances between points in different manifolds using a standard metric, which can be helpful in specific applications. 
			
			\begin{definition}[\bf Euclidean Pullback Metrics (EPM)]
				\label{def2} Let  $(\mathcal{M}_n,g^f)$ be a Riemannian Manifold. Then $g$ is Euclidean Pullback Metric if there is a diffeomorphism $f$ such that $g^f$ is the Pullback Metrics Definition (\ref{pullb}) of the Frobenius metric $g^E$ by $f$. 
			\end{definition} 
			
			If a Riemannian metric $g$ on a space $\mathcal{M}_n$ can be written as the pullback of the Euclidean metric by a diffeomorphism $f$, then both manifolds $(\mathcal{M}_n, g)$ and $(f(\mathcal{M}_n),g^E)$ are IRMs. This is because $f: \mathcal{M}_n\to f(\mathcal{M}_n)$ is a diffeomorphism, and $g$ is the pullback of the Euclidean metric $g^E$ by $f$. In practice, many metrics $g$ used to study the properties of the SPD Manifold can be expressed as Euclidean Pullback Metrics (EPMs). Lemma (\ref{lemma1}) summarizes the most important EPM on the SPD manifold. The proof is in {Appendix} (\ref{LogCholEPM}).  
			
			\begin{lemma}\label{lemma1}
				All  Riemannian metrics $g$ in Table  (\ref{Tab2}) are $\operatorname{EPMs}$, by the diffeomorphism $f:\operatorname{SPD}_n\to f(\operatorname{SPD}_n)$
			\end{lemma}
			
			\begin{table*}[h!]
				\TBL{\caption{\small \bf   Inducing  Isometries $f$ of Euclidean Pullback Metrics $g$ on the SPD Manifold}}
				{\begin{tabular*}{\textwidth}{@{\extracolsep{\fill}}llllll@{}}\toprule\small
						\label{Tab2}
						\TCH{\small \bf Metric $g$ } & \TCH{ \small$f(Y)$,\,\,\,($Y\in \operatorname{SPD}_n$)} &$f(\operatorname{SPD}_n)$& 
						\\\midrule
						\small
						\small Power - Euclidean &\small$\operatorname{pow}_\tau(Y)$   &\small$\operatorname{SPD}_n$     \\ 
						\small Log - Euclidean  &\small $\log \left(Y\right)$  & \small$\operatorname{Sym}_n$     \\ \small
						Cholesky &\small $\mathcal{L}(Y)$&\small \small$\operatorname{LT}_n$   \\ 
						\small Log - Cholesky & \small $\lfloor \mathcal{L}(Y)\rfloor +\log\left(\mathbb{D}(\mathcal{L}(Y))\right)$  &\small $\operatorname{LT}_n$       
						\\\botrule
					\end{tabular*}}
				\end{table*}
				\subsection{Taylor expansion for Multivariate Regression  }\label{tayyyyy}
				Expressing LPR as a weighted least square optimization problem requires the Taylor expansion of the regression function  \cite{ruppert1994multivariate,fan2018local,wand1994kernel}. Our interest is to study regression problems where the covariates belong to Euclidean space and the response variable to a Riemannian Matrix Manifold, which are nonlinear subspaces of $\operatorname{M}_{n,l}(\mathbb{R})$. This means that the regression function $m(\cdot)$ is a vector value matrix function, and therefore a good Taylor expansion is needed \cite{chacon2020higher,van2019taylor}. Various proposals to expand  $m({X})$ can be found in the literature due to the existence of multiple equivalent ways to define the partial derivative of a vector value matrix function with respect to a vector \cite{magnus1985matrix,magnus2010concept,van2010consistent}. Among these proposals, we chose to use the Taylor expansion given in \cite{van2019taylor} since it is a more natural and intuitive version. This expansion uses the Definition of the derivative with respect to a vector exposed in {Appendix} (\ref{deriv}). According to this definition, every vector value function $m(X)\in \operatorname{M}_{n,l}(\mathbb{R})$ that is $k+1$ times differentiable at $x$ in $\mathbb{R}^p$ has a Taylor expansion in a neighborhood of $x$
				\begin{align}\label{taylorcoef}
					m(X)-m(x) &\approx \sum_{j=1}^{k} \alpha_j\left(I_l\otimes (X-x)^{\otimes j }\right),
					\\ j!\alpha_j&=\partial^j_{x} m(X), \,\,\, j=1,2,\cdots,k,
				\end{align} 
				where $\alpha_j \in \operatorname{M}_{n,mp^j}(\mathbb{R})$ is partial derivative of order $j$ evaluated in $x$ \cite{van2019taylor}.  One important tool that we need to handle the partial derivative is { Lemma} (\ref{productderivate}). This result has been proven in \cite{van2010consistent,van2019taylor} and is very helpful in proving the main findings of this research.
				\begin{lemma}\label{productderivate}
					If $x\in\mathbb{R}^p$, $A(x)\in \mathcal{C}^{(\infty)}(\mathbb{R}^p,M_{n}(\mathbb{R}))$ and $B(x)\in \mathcal{C}^{(\infty)}(\mathbb{R}^p,M_{n}(\mathbb{R})) $ then 
					\begin{align}
						{\partial}_{ x}\left(A(x)B(x)\right)={\partial_x A(x)} \left(B(x)\otimes I_p\right)+A(x){\partial_x B(x)}. 
					\end{align}
					In particular $B(x)=B$  is constant then 
					\begin{align}
						\partial^j_x\left(A(x)B\right)=\partial^j_x A(x) \left(B\otimes I_p^{\otimes j}\right).
					\end{align}
				\end{lemma}

\section{Multivariate Intrinsic Local Polynomial Regression on Isometric Riemannian Manifolds (ILPR-IRM) } \label{metods}
\subsection{Multivariate Local Polynomial Regression on Euclidean Space (LPR) }\label{lpr}
A Riemannian matrix manifold, denoted by $(\text{M}_n(\mathbb{R}),g^E)$, can be regarded as a Euclidean space equipped with the Euclidean metric. This manifold shares several geometric properties with the classical Euclidean space $\mathbb{R}^{n^2}$ equipped with the Euclidean distance. The vec operator establishes an isomorphic relationship between them. Because of these similarities, regression methods in $\mathbb{R}^{n^2}$ can be readily extended to $\operatorname{M}_n(\mathbb{R})$. Thus, the classical Definition of multivariate local polynomial regression (LPR) in $\mathbb{R}^{n^2}$ \cite{ruppert1994multivariate,fan2018local,wand1994kernel} can be directly applied to $\operatorname{M}_n(\mathbb{R})$.

Here is the formal Definition  of classical multivariate LPR of degree $k$ in the Euclidean space $(\operatorname{M}_n(\mathbb{R}),g^E)$ at $x$, given a set of independent and identically sampled points $\mathbf{\Sigma}=\{(X_i, Y_i)\in \mathbb{R}^{p}\times \operatorname{M}_n(\mathbb{R}), i=1,2,\cdots, N\}$ from a population $({X}, {Y})$:

\begin{definition}[\bf Multivariate Local Polynomial Regression (LPR)]\label{clasical}
	Let $\mathbf{\Sigma}$ be as defined above. Classical Multivariate Local Polynomial Regression (LPR) of degree $k$ in $(\operatorname{M}_n(\mathbb{R}),g^E)$ at $x$ consists of estimates of the regression function ${m}(x)=\mathbb{E}[{Y}|X=x]$ and the derivatives $\partial^j_{x}\left({m}(X)-{m}(x)\right)$ for $j=1,2,\cdots, k$, under the assumption that ${m}$ is a $k+1$ times differentiable function at $X=x$.
\end{definition}

The assumptions of Definition  \ref{clasical} imply that in a neighborhood of $x$, the vector-valued matrix function $t(X)={m}(X)-{m}(x)$ is a $k+1$ times differentiable function. Thus, we can use the Taylor expansion introduced in Subsection \ref{taylorcoef} to obtain:

\begin{align}\label{taylorexp}
	\begin{split}
		t&:\mathbb{R}^p\to \operatorname{M}_n(\mathbb{R}),
		\
		\\t(X) & = {m}(X)-{m}(x),
		\
		\\{t}(X)&\approx \sum_{j=1}^{k}\frac{1}{j!}\partial^j_{x}{t}(X)(I_n\otimes (X-x)^{\otimes j}),
	\end{split}
\end{align}
where $\partial^j_{x}{t}(X)/j!\in \operatorname{M}_{n,np^j}(\mathbb{R})$ are the Taylor coefficients of $t(X)$. This polynomial is fitted locally by a weighted least squares regression problem:

\begin{align}\label{opty1}
	\begin{split}
		\widehat{\alpha}(x,k,h) &= \arg \min_{\alpha} N^{-1}\sum_{i=1}^{N}K_{h}(X_i-x) \left|\left|m^{[k]}_{\alpha,x}(X_i)- Y_i \right|\right|_2^2,
		\\m^{[k]}_{\alpha,x}(X)&=\alpha_0+t^{[k]}_{\alpha,x}(X),
		\\t^{[k]}_{\alpha,x}(X)&=\sum_{j=1}^{k}\alpha_j \left(I_n\otimes (X-x)^{\otimes j}\right),
	\end{split}
\end{align}
where $\alpha_0 = {m}(x)$, $\alpha_j=\partial^j_{x}{t}(X)/j! \in \operatorname{M}_{n,np^j}(\mathbb{R})$ for $j=1,2,\cdots,k$ and $K_h(\cdot)$  is a Kernel function with bandwidth $h$.  Is important to remark that the $\widehat{\alpha}(x;k,h)$  is a block matrix, more precisely 
\begin{align}
	\begin{split}
		\widehat{\alpha}(x;k,h) &= (\widehat{\alpha}_0(x;k,h),\widehat{\alpha}_1(x;k,h),\cdots, \widehat{\alpha}_k(x;,k,h)),
		\\ \widehat{\alpha}_j(x;k,h)& \in \operatorname{M}_{n,np^j}(\mathbb{R}).
	\end{split}
\end{align}

Looking at Taylor expansion {Eq}. (\ref{taylorexp})  is clear that $j!\widehat{\alpha}_j(x;k,h)$  is an estimator of $\partial^j_{x}m(X)$. For scalar covariate vectors and responses $Y_i$, the optimization problem {Eq}. (\ref{opty1}) reduces to the classical univariate LPR problem \cite{fan2018local,wand1994kernel}.

\subsection{Multivariate Intrinsic Local Polynomial regression on an arbitrary Riemannian Matrix Manifold (ILPR) }\label{milpr}

As previously outlined in the Introduction, statistical methodologies such as the LPR method may prove inadequate when working with data possessing responses on a Riemannian Manifold. The geometric structure of such a manifold necessitates a consistent approach that accounts for its nonlinearity. The Intrinsic Local Polynomial Regression (ILPR) was developed to generalize LPR to conform to the geometric structure of Riemannian value data\cite{yuan2012local}. While initially formulated with the SPD Manifold, ILPR may be readily adapted to any arbitrary Riemannian Manifold. Accordingly, we propose a revised Definition of ILPR capable of accommodating more general settings involving covariate vectors in IRMs.

\begin{definition}[\bf Intrinsic Local Polynomial Regression (ILPR)]\label{genemilpr}
	Let $\mathbf{\Sigma}=\{(X_i, Y_i)\in \mathbb{R}^{p}\times \mathcal{M}_n, i=1,2,\cdots, N\}$ be a set of independent and identically distributed samples from a population $({X},{Y})$, and let $\omega \in \mathcal{M}_n$. ILPR of degree $k$ on the Riemannian Manifold $(\mathcal{M}_n,g)$ at $(x,\omega)\in\mathbb{R}^p\times\mathcal{M}_n$ is a statistical methodology used to estimate the regression function $m(x)$ such that $\mathbb{E}[\operatorname{log}_{m(X)}Y|X=x]=O_n$ and the derivatives $\partial^j_{x}\left( \pi_{ m(x)\to \omega}\left(\operatorname{log}_{m(x)}m(X)\right)\right)$, $j=1,2,\cdots, k$, under the assumption that $m$ is a $k+1$ times differentiable function at $X=x$. Here, $\pi$ and $\operatorname{log}$ represent the Parallel Transport and Logarithm Map induced by the metric $g$.
\end{definition}
The flexibility of Definition (\ref{genemilpr}) compared to the original definition of Intrinsic Local Polynomial Regression (ILPR) described in \cite{yuan2012local} when applied to the SPD manifold is because it does not require parallel transport from the matrix identity $I_n$, which makes it applicable to a broader range of Riemannian Matrix Manifolds. This enhanced flexibility allows for examining the effects of isometries when conducting regression analysis on a data set $\mathbf{\Sigma}$, thereby simplifying the mathematical complexity of the regression problem. It is worth noting that the original and new definitions of ILPR produce the same results on the $\text{SPD}_n$ manifold when $\omega = I_n$.
\begin{figure}[h!]
	\centering
	\includegraphics[width=0.5\linewidth]{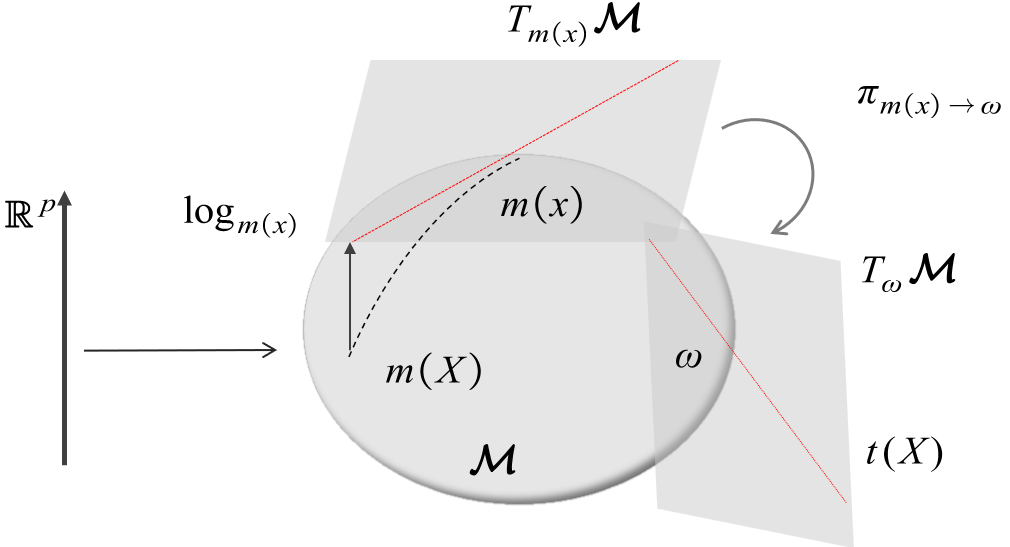}
	\caption{\small Visual representation of the function $t(X)$}
	\label{fig:Intrinsicpoly}
\end{figure}

In the case of Riemannian Matrix manifold, the function $t(X)= \pi_{ m(x)\to \omega}\left(\operatorname{log}_{m(x)}m(X)\right)$ has an image on the tangent space ${T}_{\omega}\mathcal{M}_n$, which is Euclidean and can be identified as a vector subspace of $\operatorname{M}_n(\mathbb{R})$. Therefore, it is a vector value matrix function and admits a Taylor expansion as stated in Subsection (\ref{tayyyyy}). Consequently, the assumptions in Definition (\ref{genemilpr}) imply that, in a neighborhood of $x$, we can approximate the function $t(X)= \pi_{ m(x)\to \omega}\left(\operatorname{log}_{m(x)}m(X)\right)$ by a multivariate Taylor polynomial 
\begin{align}\label{taylorexp3}
	\begin{split}
		t&:\mathbb{R}^p\to T_{\omega}\mathcal{M}_n\subseteq \operatorname{M}_n(\mathbb{R}),
		\\t(X)&= \pi_{ m(x)\to \omega}\left(\operatorname{log}_{m(x)}m(X)\right),
		\\t(X) &\approx \sum_{j=1}^{k}\frac{1}{j!}\partial^j_{x}{t}(X)(I_n\otimes (X-x)^{\otimes j}),
	\end{split}
\end{align} 
where $\partial^j_{x}{t}(X)/j!\in\operatorname{M}_{n,np^j}(\mathbb{R})$ are Taylor coefficients of $t(X)$. Like the Euclidean case, this series expansion allows rewriting the ILPR in Definition (\ref{genemilpr}) as a  weighted regression problem  that determines the Intrinsic Local Estimator 
\begin{definition}[\bf Intrinsic Local Estimator (ILE)]
	\label{def3}Let $(\mathcal{M}_n,g)$ be a Riemannian Manifold and the 
	data set $\mathbf{\Sigma}=\{(X_i, Y_i)\in \mathbb{R}^{p}\times \mathcal{M}_n, i=1,2,\cdots, N\}$. Then the  ILPR Estimator of degree $k$ in $(\mathcal{M}_n, g)$ at $(x,\omega) \in\mathbb{R}^p\times \mathcal{M}_n$ for the data $\mathbf{\Sigma}$, $\widehat{\alpha}(x;k,h) = \operatorname{ILE}(x,\omega,k,h,\mathbf{\Sigma},g)$, is the solution of the following optimization problem:
	\begin{align}\label{asddd}
		\begin{split}
			\widehat{\alpha}(x;k,h) & =  \arg \min_{\alpha} N^{-1}\sum_{i=1}^{N}K_{h}(X_i-x)\operatorname{dist}_g^2(m^{[k]}_{\alpha,x}(X_i),Y_i ) 
			\\m^{[k]}_{\alpha,x}(X)&=\operatorname{exp}_{\alpha_0}\left(\pi_{\omega\to\alpha_0}\left(t_{\alpha,x}^{[k]}(X)\right)\right),
			\\t^{[k]}_{\alpha,x}(X)&=\sum_{j=1}^{k}\alpha_j \left(I_n\otimes (X_i-x)^{\otimes j}\right).
		\end{split}
	\end{align}
	Where $K_h(\cdot)$ is a kernel function with bandwidth $h$ and $\operatorname{dist}_g$, $\operatorname{exp}$ and $\pi$ are, respectively, the Geodesic Distance, Exponential Map, and Parallel Transport induced by the metric $g$. We denote $\widehat{\alpha}(x;k,h)=(\widehat{\alpha}_0(x;k,h),\widehat{\alpha}_1(x;k,h),\cdots,\widehat{\alpha}_n(x;k,h))$, then $!\widehat{\alpha}_j(x;k,h)$ is an estimator of $\partial^j_{x}t(X) \in \operatorname{M}_{n,np^j}(\mathbb{R})$  for $j=1,2,\cdots, k$ and $\widehat{\alpha}_0(x;k,h)$ is an estimator of $m(x)$.
\end{definition}
In the Definition (\ref{def3}), we denote $\widehat{\alpha}(x;k,h)=\operatorname{ILE}(x,\omega,k,h,\mathbf{\Sigma}, g)$, for simplicity in the notation. However, each time we use $\widehat{\alpha}$ during this paper, said value is a function of  $x,\omega,k,h,\mathbf{\Sigma}$ and $g$. 

This paper uses the Leave One Out Cross Validation (LOOCV) method for bandwidth selection \cite{ruppert1994multivariate,fan2018local,wand1994kernel,avery2013literature}. The optimal bandwidth is the minimum of the LOOCV score $CV(h)$
\begin{align}\label{cross}
	\begin{split}
		\widehat{h} &= \arg\min_{h} CV(h), 
		\\CV(h) &= N^{-1} \sum_{i=1}^{N} \operatorname{dist}_g^{2}\left(\widehat{\alpha}_0^{(-i)}(X_i;k,h),Y_i\right),
		\\\widehat{\alpha}^{(-i)}(X_i;k,h)& =   \operatorname{ILE}(X_i,\omega,k,h,\mathbf{\Sigma}^{(-i)},g),
		\\\mathbf{\Sigma}^{(-i)} & = \mathbf{\Sigma} - {(X_i,Y_i)}. 
	\end{split}
\end{align}

When ignoring the geometrical information of the data $\mathbf{\Sigma}$, i.e., considering the Riemannian Manifold of the responses ($\mathcal{M}_n$) as a Euclidean subspace of $\text{M}_n(\mathbb{R})$. The ILPR in  Definition (\ref{MILPR}) reduces to the LPR in Definition (\ref{lpr}). The effects of ignoring the geometrical information translate in mathematical terms to the following changes in {Eq}. (\ref{asddd}) 
\begin{align}
	\begin{split}
		g&\to g^F,
		\\\operatorname{dist}_g^2(m^{[k]}_{\alpha,x}(X_i),Y_i )&\to \left|\left|m^{[k]}_{\alpha,x}(X_i)- Y_i \right|\right|_2^2, 
		\\m^{[k]}_{\alpha,x}(X)= \operatorname{exp}_{\alpha_0}\left(\pi_{\omega\to\alpha_0}\left(t_{\alpha,x}^{[k]}(X)\right)\right)&\to m^{[k]}_{\alpha,x}(X)=\alpha_0+t^{[k]}_{\alpha,x}(X).
	\end{split}
\end{align}

\subsection{Relationship between the  ILPRs problems on IRMs}\label{relationsec}
\begin{figure}[h!]
	\small\FIG{\includegraphics[width=0.5\linewidth]{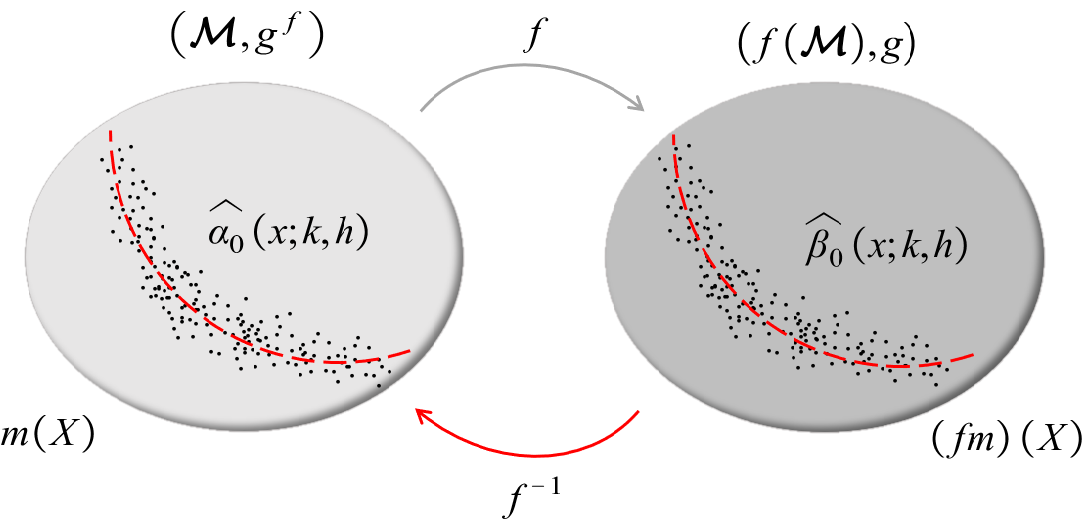}}
	{\caption{\small $(\mathcal{M},g^f)$ and $(f(\mathcal{M}),g)$ are IRMs  by the isometry $f$. The dots on the Manifolds $\mathcal{M}$ and $f(\mathcal{M})$ represents the sample $\mathbf{\Sigma}$  and $f(\mathbf{\Sigma})$  respectively. The discontinue lines represents the ILEs $\widehat{\alpha}_0(x;k,h)$ and $\widehat{\beta}_0(x;k,h)$ on their respectively Riemannian Manifold. The ILEs are related by the isometry $f$}}
	\label{chap1:fig1}
\end{figure}
In the literature, the standard approach to performing Intrinsic Linear Prediction Regression (ILPR) on a dataset $\mathbf{\Sigma}={(X_i, Y_i)\in \mathbb{R}^{p}\times \mathcal{M}_n, i=1,2,\cdots, N}$ with responses on a Riemannian Matrix Manifold $(\mathcal{M}_n, g)$ involves directly computing the ILE estimator $\widehat{\alpha}(x;k,h) = \text{ILE}(x,\omega,h,k,\mathbf{\Sigma}, g)$ without reference to an additional manifold. This involves solving a weighted least square optimization problem (Definition (\ref{def3})), which depends on the geometry of the Riemannian Manifold $(\mathcal{M}_n,g)$. However, if the geometrical structure induced by the Riemannian metric $g^f$ is complicated, a direct search for the solution may not be feasible without supplementary information, and numerical methods become necessary. This approach may increase computational costs and compromise the accuracy of ILPR when compared to its extrinsic equivalent.

In practical applications, additional information about the geometric structure of the Riemannian Manifold is often available. This can help us find an isometry $f$ that maps a complicated Riemannian Manifold $(\mathcal{M}_n, g^f)$ onto a simpler one $(f(\mathcal{M}_n),g)$. To simplify the ILPR approach, we propose a new methodology called ILPR-IRMs, which leverages the additional information contained in IRMs, specifically the isometry $f$.

Consider a dataset $\mathbf{\Sigma}$ with responses on a Riemannian Manifold $(\mathcal{M}_n, g^f)$ that requires analysis using ILPR. If we assume that $(\mathcal{M}_n,g^f)$ and $(f(\mathcal{M}_n),g)$ are IRMs through the isometry $f$, indicating that $\mathcal{M}_n$ is similar to $f(\mathcal{M}_n)$ and that $g^f$ is a metric derived from $g$ via $f$, this has important implications.

First, the isometry $f$ shifts the spaces of the responses of the data from $\mathcal{M}_n$ to $f(\mathcal{M}_n)$, inducing a new data $f(\mathbf{\Sigma})=\{(X_i, f(Y_i))\in \mathbb{R}^{p}\times f(\mathcal{M}_n), i=1,2,\cdots, N\}$ (see Figure (\ref{chap1:fig1})). Second, determining the geometrical structure using the Riemannian Metric with IRMs-isometry $f$ is more relaxed and easier to analyze.

Therefore, the use of IRMs-isometry $f$ can simplify the dataset analysis by enabling us to perform a second ILPR problem on the induced data $f(\mathbf{\Sigma})$ with responses on the Riemannian Manifold $(f(\mathcal{M}_n),g)$. This new problem is equivalent to the original but has a more straightforward geometrical structure.

\begin{align*}
	\begin{matrix}
		\text{ILPR on } (\mathcal{M}_n,g^f)
		\\ \text{------------------------------}  
		\\ \text{Data } 
		\\ \mathbf{\Sigma}=\{(X_i, Y_i)\in \mathbb{R}^{p}\times \mathcal{M}_n, i=1,2,\cdots, N\}
		\\ \text{------------------------------}  
		\\ \text{Intrinsic Local Estimator}
		\\\,\,\,\,\,\widehat{\alpha}(x;k,h)=\operatorname{ILE}(x,\omega,k,h,\mathbf{\Sigma},g^f)   \\ 
	\end{matrix} &\underrightarrow{\text{\,\,\,\,\,\,\,\,\,\,\,f\,\,\,\,\,\,\,\,\,\,} } \begin{matrix}
	\text{Induced ILPR on } (f(\mathcal{M}_n),g)
	\\ \text{------------------------------}  
	\\ \text{Induced Data } 
	\\ f(\mathbf{\Sigma})=\{(X_i, f(S_i))\in \mathbb{R}^{p}\times f(\mathcal{M}_n), i=1,2,\cdots, N\}
	\\ \text{------------------------------} 
	\\ \text{Induced Intrinsic Local estimator}
	\\\widehat{\beta}(x;k,h)=\operatorname{ILE}(x,f(\omega),k,h,f(\mathbf{\Sigma}),g)
	\\
\end{matrix} 
\\ 
\end{align*}

In this context, it is unclear whether there is a relationship between $\widehat{\alpha}(x;k,h)$ and $\widehat{\beta}(x;k,h)$. This question is addressed in this section, and the main result, presented in Theorem  (\ref{MILPR}), provides a positive answer under specific hypotheses. The proof of this Theorem is in Appendix (\ref{appendixA}) and relies on Lemma  (\ref{productderivate}) and the subsequent Lemma.
\begin{lemma}\label{Lemmaiso}
	Let $L: \mathcal{A}\subseteq \operatorname{M}_n(\mathbb{R})\to \operatorname{M}_n(\mathbb{R})$ be an invertible linear map. Then there exist two invertible matrices $A,B\in \operatorname{GL}_n(\mathbb{R})$ such that for all $Z\in \mathcal{A}$, $L(Z) = AZB$.
\end{lemma}

The Lemma provides essential insight into the structure of invertible linear maps on a subset of $n\times n$ real matrices. It states that any linear map $L$ can be expressed using two invertible matrices $A$ and $B$ as $L(Z) = AZB$ for all $Z$ in the subset $\mathcal{A}$. The proof of Lemma  (\ref{Lemmaiso}) is presented in the Appendix (\ref{Lemmaproof}).
\begin{theorem}[Relationship between  ILPRs on IRMs]
	\label{MILPR}
	Under the following conditions:
	\begin{enumerate}
		\item[(a)] $(\mathcal{M}_n,g^f)$ and $(f(\mathcal{M}_n),g)$ are Isometric Riemannian Manifolds (IRMs) related by the isometry $f$.
		\item[(b)] For all $(x,\omega)\in\mathbb{R}^p\times\mathcal{M}_n$, the Intrinsic Local Estimator (ILE) maps $\widehat{\alpha}(x;k,h)=\operatorname{ILE}(x,\omega,k,h,\mathbf{\Sigma},g^f)$ and $\widehat{\beta}(x;k,h)=\operatorname{ILE}(x,f(\omega),k,h,f(\mathbf{\Sigma}),g)$ exist.
		\item[(c)] There exist unique matrices $A,B\in\operatorname{GL}_n(\mathbb{R})$ such that $d_{f(\omega)}f^{-1}(V)= AVB$ for all $V\in T_{f(\omega)}f(\mathcal{M}_n)$.
	\end{enumerate}
	The following relation holds between the ILE maps:
	\begin{equation}
		\label{polypoly}
		\widehat{\alpha}_j(x;k,h) = \delta_{j0}f^{-1}(\widehat{\beta}_j(x;k,h)) + (1-\delta_{j0})A\widehat{\beta}_j(x;k,h)(B\otimes I_p^{\otimes j}), \quad \text{for } j=0,1,2,\dots,k,
	\end{equation}
	where $\delta_{j0}$ is the Kronecker delta and $I_p^{\otimes j}$ is the $j$-fold tensor product of the $p\times p$ identity matrix.
\end{theorem}

Theorem (\ref{MILPR}) is a valuable tool in solving the ILPR problem on IRMs because it enables the transfer of the ILPR problem from one Riemannian Manifold to another, ultimately solving it in the most convenient one, as shown in Figure (\ref{chap1:fig1}). This article will utilize this Theorem to derive a closed analytical expression for the ILE in Riemannian Manifold equipped with an EPM.

\subsection{Exact solution of the ILPR-EPM } \label{euclsec}
In this section, we have obtained the precise analytical form of the multivariate ILPR on a manifold $\mathcal{M}_n$ with a metric $g^f$ belonging to the Euclidean Pullback Metric (EPM) family, as defined in {Eq}. (\ref{def2}). The key outcome of this section is Theorem (\ref{teorema3}), which is a consequence of Theorem (\ref{MILPR}).

\begin{theorem}[ Exact analytical solution of the ILPR-EPM]\label{teorema3}
	Under the following  conditions  
	
	\begin{enumerate}
		\item[(a)]  $(\mathcal{M}_n,g^f)$  is a Riemannian Manifold equipped with a Euclidean Pullback Metric (EPM) $g^f$ induced by the isometry $f$.
		\item[(b)] For all  $(x,\omega)\in\mathbb{R}^p\times\mathcal{M}_n$, the Intrinsic Local Estimators (ILEs) $\widehat{\alpha}(x;k,h)=\operatorname{ILE}(x,\omega,k,h,\mathbf{\Sigma},g^f)$ and $\widehat{\beta}(x;k,h)=\operatorname{ILE}(x,f(\omega),k,h,f(\mathbf{\Sigma}),g^E)$ exist.  	
	\end{enumerate}
	Then the closed analytical expression of the $\widehat{\alpha}_0(x;k,h)$ and $\widehat{\beta}_0(x;k,h)$ are:  
	\begin{align}
		\label{s1}  
		\widehat{\alpha}_0(x;k,h) &= f^{-1}\left(\widehat{\beta}_0(x;k,h)\right),
		\\\label{s2} \widehat{\beta}_0(x;k,h)& = f(\mathbf{Y})\left((\mathbf{WX}^T(\mathbf{XWX}^T)^{-}e_0)\otimes I_n\right),
	\end{align}
	where 
	\begin{align}\label{s3}
		\begin{split}
			e_0 &= (1, 0,\cdots,0)^T \in \operatorname{M}_{\sum_{j=0}^{k}p^j,1}(\mathbb{R}),
			\\f(\mathbf{Y})&=(f(Y_1), f(Y_2),\cdots,f(Y_N))\in \operatorname{M}_{n,nN}(\mathbb{R}),
			\\\mathbf{X} &=\begin{bmatrix}
				1 & 1& \cdots& \cdots &1
				\\(X_1-x)^{\otimes 1}& (X_2-x)^{\otimes 1} &\cdots& \cdots &(X_N-x)^{\otimes 1}
				\\(X_1-x)^{\otimes 2}& (X_2-x)^{\otimes 2}& \cdots& \cdots &(X_N-x)^{\otimes 2}
				\\\cdots& \cdots& \cdots &\cdots& \cdots 
				\\\cdots& \cdots& \cdots &\cdots& \cdots 
				\\(X_1-x)^{\otimes j}& (X_2-x)^{\otimes j}& \cdots& \cdots &(X_N-x)^{\otimes j}
				\\\cdots& \cdots& \cdots &\cdots& \cdots 
				\\\cdots& \cdots& \cdots &\cdots& \cdots 
				\\(X_1-x)^{\otimes k}& (X_2-x)^{\otimes k}&\cdots& \cdots &(X_N-x)^{\otimes k}
			\end{bmatrix} \in \operatorname{M}_{\sum_{j=0}^{k}p^j,N}(\mathbb{R}),
			\\\mathbf{W}&=\operatorname{diag}(K_h(X_1-x),K_h(X_2-x),\cdots,K_h(X_N-x))\in \operatorname{M}_{N,N}(\mathbb{R}).
		\end{split}
	\end{align}
\end{theorem}
The proof of {Theorem} (\ref{teorema3}) is elaborated in  {Appendix} (\ref{Proofteorem3}).  An important observation from Eqs. (\ref{s1},\ref{s2}) in  Theorem (\ref{teorema3})  is that
\begin{align}
	\widehat{\alpha}_0(x;k,h) = f^{-1}\left( f(\mathbf{Y})\left((\mathbf{WX}^T(\mathbf{XWX}^T)^{-}e_0)\otimes I_n\right)\right).
\end{align}
This result provides the exact analytical expression for the multivariate Intrinsic Local Polynomial Regression (ILPR) on any Riemannian Manifold equipped with a metric in the Euclidean Pullback Metric (EPM) family.

As an illustrative example, spouse that is desired finds the Local Constant estimators $\widehat{\alpha}(x;0,h) =\operatorname{ILE}(x,\omega,0,h,\mathbf{\Sigma},g^f)$ where $g^f$ is an EPM by the isometry $f$. Theorem (\ref{teorema3}) states  that in order to compute $\widehat{\alpha}_0(x;0,h)$ we can compute $\widehat{\beta}_0(x;0,h)$, which is the induced local estimator by the isometry $f$. Then Theorem (\ref{teorema3}) states that such estimators have a closed analytical expression 
determined by  the formula  Eq. (\ref{s2}) where  $f(\mathbf{Y})$ and $\mathbf{W}$ is defined as in Eq. (\ref{s3}) and $e_0$ and $\mathbf{X}$ in the local constant regression  the particular expression 
\begin{align}
	\begin{split}
		e_0  =  1,
		&\,\,\, \mathbf{X} = 1_{N\times 1}^T,
	\end{split}
\end{align}
from this expression can compute $\mathbf{WX}^T(\mathbf{XWX}^T)^{-}e_0$ as 
\begin{align*}
	\mathbf{WX}^T(\mathbf{XWX}^T)^{-}e_0 &= \mathbf{W} 1_{N\times 1}(1_{N\times 1}^T\mathbf{W} 1_{N\times 1})^{-}1
	\\& = \frac{(K_h(X_1-x),\cdots,K_h(X_N-x))^T}{\sum_{i=1}^{N}K_h(X_i-x)},
\end{align*}
now we can compute the local constant estimator $\widehat{\beta}_0(x;0,h)$ in the following manner 
\begin{align}\label{examplebeta}
	\begin{split}
		\widehat{\beta}_0(x;0,h)& = f(\mathbf{Y})\left((\mathbf{WX}^T(\mathbf{XWX}^T)^{-}e_0)\otimes I_n\right)
		\\& =  (f(Y_1), f(Y_2),\cdots,f(Y_N))\left(\frac{(K_h(X_1-x),\cdots,K_h(X_N-x))^T}{\sum_{i=1}^{N}K_h(X_i-x)}\otimes I_n\right)
		\\& = (f(Y_1), f(Y_2),\cdots,f(Y_N))\frac{(K_h(X_1-x)I_n,\cdots,K_h(X_N-x)I_n)^T}{\sum_{i=1}^{N}K_h(X_i-x)}
		\\& = \frac{\sum_{i=1}^{N}K_h(X_i-x)f(Y_i)}{\sum_{i=1}^{N}K_h(X_i-x)}.
	\end{split}
\end{align}
Finally by Eq. (\ref{s1})  we can compute $\widehat{\alpha}(x;0,h)$ as follows  
\begin{align}\label{nadaraya}
	\begin{split}
		\widehat{\alpha}(x;0,h)& = f^{-1}(\widehat{\beta}_0(x;0,h)) \\=&f^{-1}\left(\frac{\sum_{i=1}^{N}K_h(X_i-x)f(Y_i)}{\sum_{i=1}^{N}K_h(X_i-x)}\right). 
	\end{split}
\end{align}
It is noteworthy to observe that the expression for the induced local estimator $\widehat{\beta}_0(x;0,h)$ on the Euclidean space $(f(\mathcal{M}_n),g^E)$ bears a resemblance to that of the Nadaraya-Watson estimator, as expected. Generally, the estimator $\widehat{\beta}_0(x;k,h)$ on the Euclidean space $(f(\mathcal{M}_n),g^E)$ is similar to LPR. Specifically, in the case of a univariate covariate  ($p=1$), the vectorization of Eq. (\ref{s2}) yields the exact expression for LPR in the Euclidean space.

The algorithm for computing the multivariate ILPR for a sample with responses on any Riemannian Matrix Manifold equipped with an EPM metric is illustrated in the pseudocode shown in (\ref{seudo}). Notably, the pseudocode includes a Tikhonov-Regularization procedure that can enhance the numerical stability of the regression process. A MATLAB implementation of the complete algorithm is available on GitHub, accessed via the following URL: \href{URL}{https://github.com/ronald1129/Multivariate-Intrinsic-Local-Polynomial-Regression-on-Isometric-Riemannian-Manifolds.git}.

\begin{algorithm}[h!]
	\small
	\label{seudo}
	\caption{ \small \textit{Intrinsic Local Polynomial with  Euclidean Pullback Metrics ({\bf ILPR-EPM)}}}
	\begin{algorithmic}[1]
		
		\Require{$\mathbf{\Sigma}= \{ (X_i, Y_i) \in \mathbb{R}^p\times \mathcal{M}_n, i= 1,2, \cdots, N \}$,$ x\in\mathbb{R}^p$,$k$, $h$ , $K_h(\cdot)$, $f$} 
		\Ensure{$\widehat{\alpha}_0(x;k,h)$  }
		\State $\lambda\leftarrow 10^{-3}$,  Tikhonov-Regularization parameter
		\State $\text{pdim} \leftarrow \sum_{j=0}^{k}p^j$ 
		\State $f(\mathbf{Y})\, \leftarrow  [\,\, ]$
		\State $\mathbf{X}\,\leftarrow  [\,\, ]$
		\State $\mathbf{W} \leftarrow O_N$
		\For{$i=1:N$}
		\State $f(\mathbf{Y})\, \leftarrow \left[f(\mathbf{Y}),\, f(Y_i)\right]$
		\State $\mathbf{W}(i,i)\leftarrow K_h(X_i-x)$
		\State$\text{Xcol}_i \leftarrow [ 1 ]$
		\For{$j=2:k+1$}
		\State $\text{Xcol}_i\leftarrow \left[\text{Xcol} _i; \, (X_i-x)^{\otimes(j-1)} \right]$
		\EndFor
		\State \textbf{end}
		\State $\mathbf{X}\leftarrow \left[\mathbf{X}, \, \text{Xcol}_i\right]$
		\EndFor
		\State \textbf{end}
		\State $ \widehat{\beta}_0(x;k,h) \leftarrow f(\mathbf{Y})\left((\mathbf{WX}^T(\mathbf{XWX}^T+ \lambda^2I_{\text{pdim}})^{-}e_0)\otimes I_n\right) $
		\State $\widehat{\alpha}_0(x;k,h) \leftarrow f^{-1}(\widehat{\beta}_0(x;k,h))$
		\State{\textbf{return} $\widehat{\alpha}_0(x;k,h)$}
	\end{algorithmic}
\end{algorithm}

The SPD manifold equipped with an EPM is a specific case covered by Theorem (\ref{teorema3}). Hence, we can effectively leverage the Theorem's closed analytical formula to calculate the ILPR for EPMs on SPD. This is particularly beneficial since the Theorem provides the exact analytical expression for infinitely many metrics in the EPM family, including essential metrics like Log-Cholesky, Log-Euclidean, Power-Euclidean, and Cholesky, which are discussed in Subsection (\ref{DeformedRiemannianManifold}) and known to be EPMs, as established by Lemma (\ref{lemma1}).

In terms of the asymptotic properties of the estimator $\widehat{\alpha}_0(x;k,h)$ on the Riemannian Manifold $(\mathcal{M}_n,g^f)$, it is beneficial to utilize the similarities between the estimator $\widehat{\beta}_0(x;k,h)$ on the Euclidean space $(f(\mathcal{M}_n),g^E)$ and the LPR estimator. Building on this, we establish Theorem (\ref{cons}), which confirms the consistency of the local constant and local linear ILPR for an EPM, specifically $\widehat{\alpha}_0(x;k,h)$ for $k=0,1$. Theorem (\ref{cons}), which is elaborated in Appendix (\ref{teoconsis}), provides an exact formula for the asymptotic bias of the estimator $\widehat{\beta}_0(x;1,h)$. This formula implies the consistency of the estimator $\widehat{\alpha}_0(x;1,h)$, which is a desirable property for statistical inference.

\begin{theorem}\label{cons}
	Let $x$ be a fixed element in the interior of $\operatorname{supp}(F_X)$. Assume that the Regularity Conditions $(R1) - (R3)$ holds. Then 
	\begin{enumerate}
		\item [(a)] $
		\mathbb{E}\left[\widehat{\beta}_0(x;1,h) - (fm)(x)|X_1,\cdots,X_N\right] =\frac{1}{2}h^2 \mu_2(K) \partial^2_x(fm)(X)\operatorname{vec}(I_p)\otimes I_n +o_p(h^2 I_n).$
		\item [(b)] $\mathbb{E}\left[\widehat{\alpha}_0(x;1,h) - m(x)|X_1,\cdots,X_N\right]  \to_{N\to\infty} O_n$.
	\end{enumerate}
\end{theorem}

\section{Simulations on the SPD Manifold} \label{simulations}

On the SPD manifolds, Extrinsic Local Polynomial Regression (LPR) with the Affine-Invariant metric is widely used to conduct non-parametric regression \cite{barachant2010riemannian,li2022harmonized,lin2019extrinsic,sabbagh2020predictive}. This methodology involves selecting a reference point on the manifold, such as the intrinsic mean (also known as the Karcher mean) or the identity matrix $I_n$, and then locally embedding the data into a high-dimensional Euclidean space based on that reference point. The Affine-Invariant metric is commonly utilized due to its congruence invariance property, which has numerous practical applications in fields like Brain-Computer Interfaces, Computer Vision, and Machine Learning \cite{barachant2010riemannian,barachant2013classification}. However, non-parametric inference using this metric can be computationally expensive, necessitating excessive logarithmic calculations \cite{lin2019riemannian}. Furthermore, this approach's extrinsic nature is inappropriate for highly distorted data, which refers to data where the elements are significantly far from the reference point, making it challenging to embed them accurately in a high-dimensional Euclidean space that reflects the underlying geometry of the manifold.

The earlier theoretical results, Subsection (\ref{euclsec}), establish that equipping a Manifold with EPM enables the computation of the ILPR estimator using a closed analytical expression. This study applies this general theory to perform ILPR on the SPD manifold, explicitly using the Log-Cholesky and Log-Euclidean metrics, which are EPMs. These metrics require less computation of matrix logarithms, and therefore, they are more computationally efficient than the Affine-Invariant metric but lack general congruence invariance \cite{lin2019riemannian,dryden2009non}. We compared the ILPR estimator with the Log-Cholesky and Log-Euclidean metrics against extrinsic LPR using the Affine-Invariant metric in simulations to evaluate their performance. This latter metric is a standard method in the literature. We aimed to assess the trade-off between regression error and computation time for each method while varying the manifold and covariate dimensions.

To evaluate the performance of each regression method, we conducted a Monte Carlo simulation where the manifold and covariate dimensions were restricted to $(p,n) \in ([1,6] \times [3,18]) \cap \mathbb{N}^2$. For each simulation given a covariate and manifold dimension $(p,n)$, we generated 100 realizations of simulated regression problem with TRUE DATA, DATA + NOISE, ESTIMATED DATA with responses in the manifold $\operatorname{SPD}(n)$ and a covariate in $\mathbb{R}^p$. We calculated the Root Square Mean Error (RMSE) and the Computational Time (CPU-Time) for each realization and regression method: ILPR Log-Cholesky, ILPR Log-Euclidean, and LPR Affine-Invariant.

To generate the {TRUE DATA} responses  the following function is used $m(X,p,n)$
\begin{align}\label{truemodel}
	\begin{split}
		\left(\log m(X,p,n)\right)_{ij} &=  \delta_{ij}\frac{{1}_p^TX}{2}+(1-\delta_{ij})\frac{{\lambda}^T_{ij}X}{||{\lambda_{ij}}||},\,\,\, i,j =1,2,\cdots,n
		\\{\lambda}_{ij} & ={\lambda}_{ji}, \,\,\, {\lambda}_{ij} = \operatorname{U}_p(0,1).
	\end{split}
\end{align}
where $\delta_{ij}$, is the Kronecker delta and  the vector ${1}_p \in \mathbb{R}^p$, has all components equal to one. The covariate vectors are generated using the model $X_i \sim \mathcal{N}_p({0}_{p\times 1}, I_p) \in \mathbb{R}^p$ for $i=1,\cdots, N$, where $N$ is the sample dimension per simulation. Therefore, the TRUE DATA in a simulation with parameter $(p,n)$ is given by $\mathbf{\Sigma}^{TRUE}_{p,n} = \{(X_i, m(X_i,p,n)\in\mathbb{R}^p\times\operatorname{SPD}_n: i=1,\cdots, N\}$. Here, $m(X_i,p,n)$ denotes the true covariance matrix corresponding to the covariate $X_i$ of dimension $p$ and manifold dimension $n$.

To obtain the DATA + NOISE dataset, the responses of the true data $\mathbf{\Sigma}^{TRUE}_{p,n}$ are contaminated with noise using the Riemannian Log-Normal noise model \cite{yuan2012local,schwartzman2006riemannian}. Other approaches to contaminating a sample on the SPD manifold with noise, such as the Log-Normal and Rician noise models, have been proposed in the literature \cite{yuan2012local,Zhu2007b}. For the sake of simplicity, this paper only uses the Riemannian Log-Normal model. However, the procedure presented here can be generalized to any arbitrary noise model, an exciting topic for further research. The Riemannian Log-Normal model is defined as follows:
\begin{align}\label{noisemodel}
	\begin{split}
		m^{NOISE}(X_i,p,n)& = L\exp(\mathcal{E}_i)L^T,
		\\L&=\mathcal{L}(m(X_i,p,n)),
		\\\mathcal{E}_i& = \operatorname{vech}^{-1}\left(\mathcal{N}_{n(n+1)/2}(0_{(n(n+1)/2)\times 1}, (0.5)^2I_{n(n+1)/2})\right),
	\end{split}
\end{align}
where $\operatorname{vech}^{-1}$ is the inverse half vectorization operator \cite{koning1991block}. Therefore the DATA + NOISE  is $\mathbf{\Sigma}_{p,n}^{NOISE}=\{(X_i, m^{NOISE}(X_i,p,n)\in\mathbb{R}^p\times\operatorname{SPD}_n:i=1,\cdots,N\}$. 

As previously mentioned, the ESTIMATED DATA is obtained by estimating from the noisy data $\mathbf{\Sigma}_{p,n}^{NOISE}$ using only the Local Linear case for three different methods: ILPR Log-Cholesky, ILPR Log-Euclidean, and extrinsic LPR Affine-Invariant. To do so, the ILPR estimator for the Log-Cholesky and Log-Euclidean metric is computed and evaluated at $X_i$ using Algorithm (\ref{seudo}) for the corresponding isometry $f$ of each metric, as shown in Table (\ref{Tab2}).
\begin{align}
	\begin{split}
		f^{\text{LC}}(S)& = \lfloor\mathcal{L}(S)\rfloor + \log \mathbb{D}(\mathcal{L}(S)),
		\\f^{\text{LE}}(S)& = \log(S).
	\end{split}
\end{align} 
The optimal bandwidth $\widehat{h}$ is computed for each regression method using LOOCV, with Eq. (\ref{cross}), the intrinsic Riemannian distance $\operatorname{dist}_{g^f}$ generated by each considered metric and the Gaussian Kernel function \cite{fan2018local,wand1994kernel}. The simulations focus solely on each regression method's local linear case ($k=1$). Thus, the {ESTIMATED DATA} for each metric $g^f$ (Log-Cholesky, Log-Euclidean, Affine-Invariant) is obtained 
$\mathbf{\Sigma}^{ESTIM}_{p,n,G} = \{(X_i,\widehat{\alpha}_{0}(X_i;k,h))\in \mathbb{R}^p\times\operatorname{SPD}_n: i=1,\cdots,N\}$.

To evaluate the performance of each regression method in each simulation, we compute the Root Mean Squared Error (RMSE) between the {TRUE DATA} and the {ESTIMATED DATA} for each metric $g^f$. The formula for this computation is exact and given by:
\begin{align}
	\operatorname{RMSE}_{AI}^2 = \frac{1}{N} \sum_{i=1}^{N} \operatorname{dist}_{AI}^2 \left(\widehat{\alpha}_{0}(X_i;k,h), m(X_i,p,n)\right). 
\end{align}

To compute the computational time, CPU-Time, we use an internal function of Matlab to measure each regression method's time to calculate the {ESTIMATED DATA}.

The results of the Monte Carlo simulation are presented in two independent subsections. First Subsection (\ref{3}) focuses on the manifold SPD(3) results, which is the manifold of Diffusion Tensor fields in DTI \cite{arsigny2006log,pennec2006riemannian}. In this case, the regression problem will be represented using the Ellipsoidal representation of the data \cite{westin2002processing,masutani2003mr}. Meanwhile, Subsection (\ref{high}) displays the quantitative and qualitative behavior of the different regression methods when varying the manifold and covariate dimensions.

The commonly used Ellipsoidal method in DTI is no longer sufficient to represent high-dimensional (SPD($n$), $n>>3$) data. To address this issue, a recent paper proposes using two complementary dimensionality reduction techniques, Rie-SNE and Linearized PGA, for visualizing high-dimensional regression problems \cite{bergsson2022riesne,fletcher2004principal}. 

Rie-SNE is an advanced version of the popular t-SNE visualization technique that can work on any arbitrary manifold with a Riemannian metric \cite{bergsson2022riesne,van2008visualizing,wattenberg2016use}. This method allows for the non-linear embedding of data into a lower-dimensional space while preserving the "neighborhood" condition and respecting the underlying geometry of the manifold. To determine if two elements belong to the same neighborhood, Rie-SNE requires the computation of the Riemannian distance between them and the determinant of the metric tensor for embedding. However, these computations can be computationally demanding tasks that rapidly scale with the number of data elements  \cite{bergsson2022riesne}. A fast and efficient version of Rie-SNE is necessary for big data sets, making selecting the appropriate Riemannian metric an essential step before performing Rie-SNE. In this paper, we propose using the Log-Cholesky metric on the SPD manifold to achieve the desired efficiency of Rie-SNE for this manifold. This metric does not require massive computation of matrix logarithms. It behaves similarly to the Log-Euclidean and Affine-Invariant metrics in other problems, making it suitable for large datasets \cite{lin2019riemannian}. We refer to this technique as Log-Cholesky Rie t-SNE in this paper.     

Linearized PGA is another known technique that extends classical PCA \cite{ringner2008principal,karamizadeh2013overview} from Euclidean space to the Riemannian Manifold to visualize Riemannian-valued data \cite{fletcher2004principal}. It extracts the most informative directions in the manifold that describe the variance in the data, making it a valuable tool for visualizing the structure and variability of high-dimensional data. Linearized PGA is used in the paper to compare the regression results obtained using Log-Cholesky Rie t-SNE. 

The paper expects a suitable regression method to distribute the ESTIMATED DATA close to the TRUE DATA in both Log-Cholesky Rie t-SNE and Linearized PGA visualization techniques. Therefore, the DATA+NOISE is isolated, and the regression method can be considered successful in capturing the underlying structure of the high-dimensional data.

\subsection{Simulation I. DTI Type Data }\label{3}
\begin{figure}[h!]
	\centering
	\includegraphics[width=0.7\linewidth]{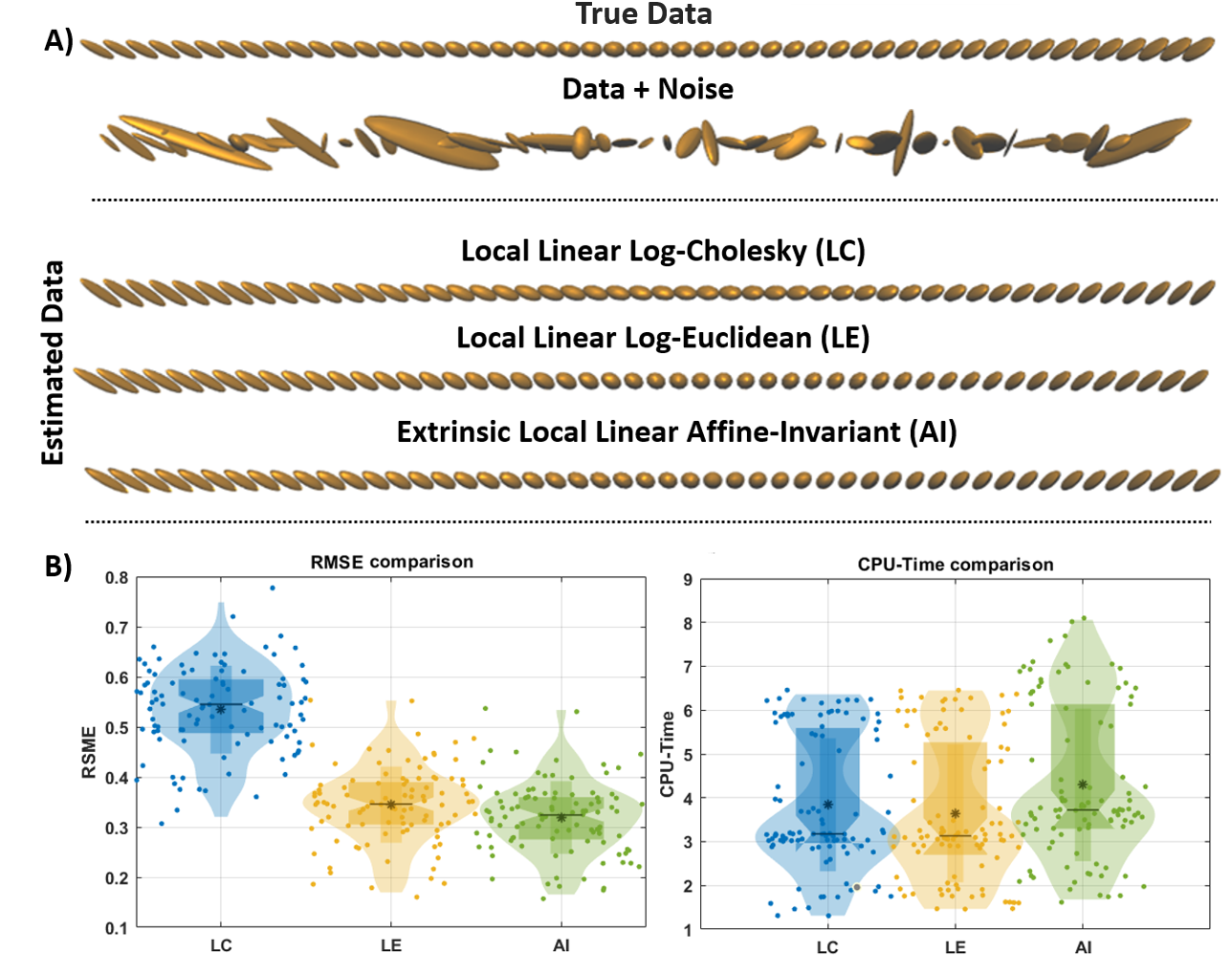}
	\caption{\small Figure compares the performance of three non-parametric regression methods on SPD(3) manifold using Monte Carlo simulation. {\bf A)} The Ellipsoidal Representation shows a single realization of TRUE DATA and DATA+NOISE. All the methods have similar performance in dealing with noise. {\bf B)} Figure compares the RMSE and CPU-Time for each method using boxplots. For DTI simulated data, the Log-Euclidean method provides the best trade-off between error and time for performing nonlinear regression.}
	\label{fig:trade_off1}
\end{figure}

This section of the paper studies the performance of non-parametric regression methods developed compared to those established in the literature on the SPD manifold of dimension 3. As mentioned in the previous section, this manifold arises in DTI analysis. In DTI a symmetric positive definite matrix represents the diffusion of water molecules corresponding to each voxel. To assess the performance of the non-parametric regression methods, we conducted a Monte Carlo simulation on the manifold SPD(3) for a scalar covariate. Specifically, we generated $N=100$ realizations of TRUE DATA, which we then used to compute the RMSE and CPU-Time for each non-parametric regression method, as described in the previous section.

In Figure (\ref{fig:trade_off1}. A), we present a visualization of a single realization of the TRUE DATA and DATA+NOISE using an Ellipsoidal Representation. The TRUE DATA was generated from the model Eq. (\ref{truemodel}) for $n=3$ and $p=1$, while DATA+NOISE was generated from the TRUE DATA using the noise model Eq. (\ref{noisemodel}). The following three rows correspond to the ESTIMATED DATA using the three non-parametric regression models discussed at the beginning of this section: intrinsic local linear Log-Cholesky and Log-Euclidean, and extrinsic local linear Affine-Invariant.

As shown in Figure (\ref{fig:trade_off1}. A), the ESTIMATED DATA from each regression method visually resembles the TRUE DATA more than the DATA+NOISE. All the regression methods capture the complex nonlinear pattern and effectively remove the noise from the DATA+NOISE, indicating that they are well-suited to this type of data.

Figure (\ref{fig:trade_off1}. B) provides further insight into the performance of the regression methods. Here, we compare the RMSE and CPU-Time for each method using boxplots. We observe that the Local Linear Log-Cholesky estimator has the highest RMSE, while the extrinsic method using Affine-Invariant has the lowest. Interestingly, the Log-Euclidean estimator performs similarly to the Affine-Invariant estimator.

Furthermore, Figure (\ref{fig:trade_off1}. B) also shows the computational time required for each method to compute the ESTIMATED DATA, CPU-Time. We observe that the intrinsic local linear Log-Euclidean method requires the least computational time, while the extrinsic local linear Affine-Invariant method requires the most. Among all the methods for performing nonlinear regression, the Log-Euclidean method provides the best trade-off between error and time. It performs relatively well while requiring fewer computational resources than other methods.
\subsection{Simulation II. High dimensional Data }\label{high}
\begin{figure}[h!]
	\centering
	\includegraphics[width=1.03\linewidth]{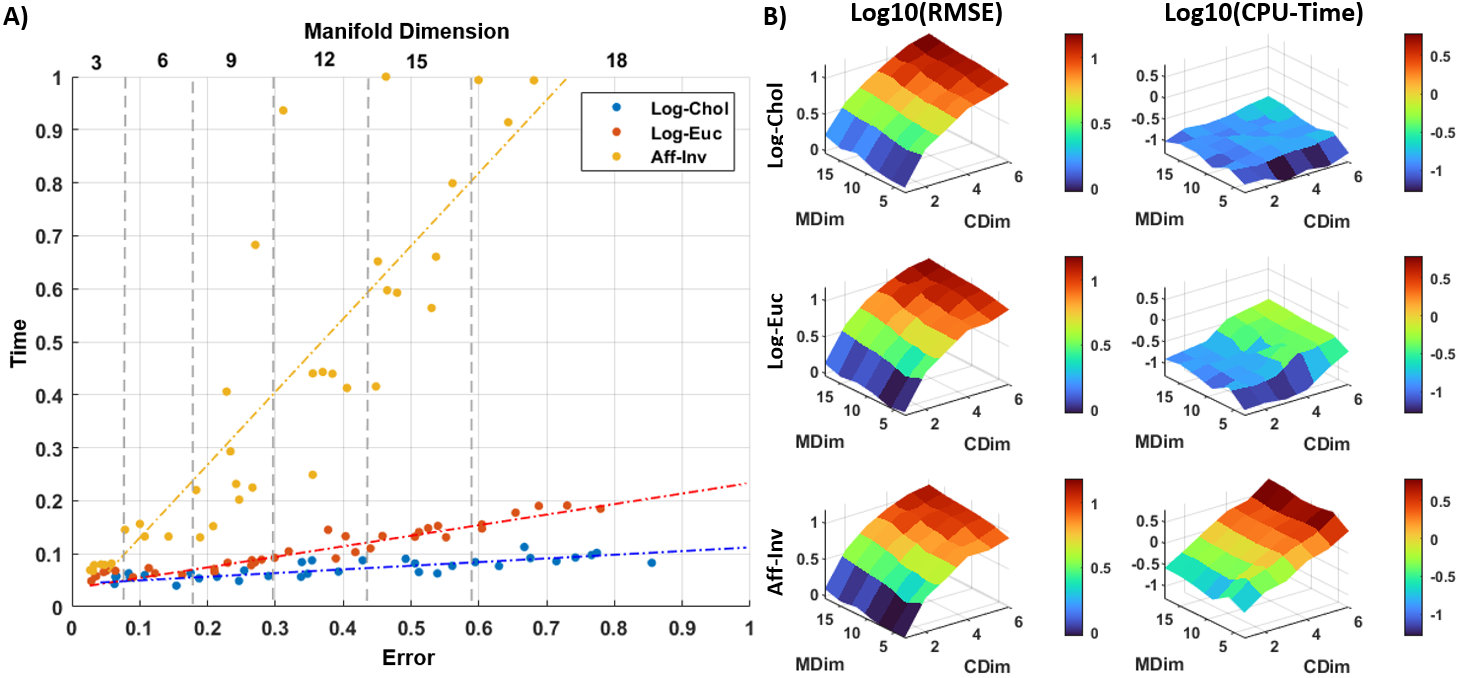}
	\caption{\small Scatter plot ({\bf A}) and heatmap ({\bf B}) illustrate the performance trade-off of non-parametric regression methods as manifold and covariate dimensions increase. {\bf A)} The normalized CPU-Time and RMSE values for intrinsic local linear estimators using Log-Cholesky and Log-Euclidean metrics on the SPD Manifold, and extrinsic local linear estimators using the Affine-Invariant metric are represented by blue, red, and yellow dots, respectively. {\bf B)} The first column of the heatmap displays the RMSE values in logarithmic scale, while the second column shows the CPU-Time values also in logarithmic scale}
	\label{fig:trade_off}
\end{figure}

In this section, we investigate the performance of the non-parametric regression methods as we increase both the manifold and covariate dimensions. To achieve this, we conducted a Monte Carlo simulation for each combination of manifold and covariate dimensions in the range $(p,n) \in ([1,6] \times [3,18]) \cap \mathbb{N}^2$.

Figure (\ref{fig:trade_off}. A) presents a scatter plot of the computed performance of all the non-parametric regression methods considered in the Monte Carlo simulation. The blue and red dots correspond to intrinsic local linear estimators using the Log-Cholesky and Log-Euclidean metrics on the SPD manifold. The yellow dots correspond to the extrinsic local linear estimator using the Affine-Invariant metric. Each dot represents the normalized value of the CPU-Time and RMSE for all methods for a specific value of the covariate dimension. The range of the covariate is at the top of the graph. We normalize the features CPU-Time and RMSE  by computing the ratio between the value after performing the Monte Carlo simulation and the maximum value of all regression methods of each feature. From Figure (\ref{fig:trade_off}. A), it is possible to appreciate that increasing the manifold dimension increases the computation time (i.e., CPU-Time) for all regression methods. In particular, the extrinsic Affine-Invariant approach is the most computationally expensive. In contrast, the computation time for the intrinsic Log-Cholesky approach is the slowest method to scale with the manifold.

Looking at the error, i.e., the RMSE, we can conclude that all the regression methods have values inside the same range. The extrinsic Affine-Invariant approach performs slightly better than the other methods. Nevertheless, the error values generally have similar values within the same range.

Figure (\ref{fig:trade_off}. B) offers an alternative representation of the data in Figure (\ref{fig:trade_off}. A). This time, the figure provides the value of each feature used to assess the performance: RMSE (first column) and CPU-Time (second column) in a logarithmic scale. From the first column, we can conclude that the RMSE for each regression method behaves qualitatively and quantitatively similarly when scaling the manifold (MDim) and covariate (Cdim) dimensions. The general tendency of the RMSE for all methods is to increase slightly with the manifold dimension and to overgrow with the number of covariates. This rapid growth with the covariate dimension could be an effect of the known phenomenon in regression literature called the "curse of dimensionality" \cite{geenens2011curse}.

The second column shows the computational time required for a Monte Carlo simulation given a manifold and covariate dimension. We can appreciate from this second column that the intrinsic Log-Cholesky approach scales very little compared to the other methods. On the other hand, the extrinsic Affine-Invariant approach to regression increases fast when scaling the manifold and covariate dimension, meaning that this is the regression method with the worst time performance. At the same time, the intrinsic Log-Euclidean approach's time performance is in the intermediate place between the other two regression methods. For all regression methods, the general tendency is for computation to increase when increasing the manifold and covariate dimension. This time performance qualitative beehive of the regression methods is an expected result related to the amount of matrix logarithm computation required for each metric \cite{lin2019riemannian,arsigny2006log}.

\begin{figure}[h!]
	\centering
	\includegraphics[width=0.9\linewidth]{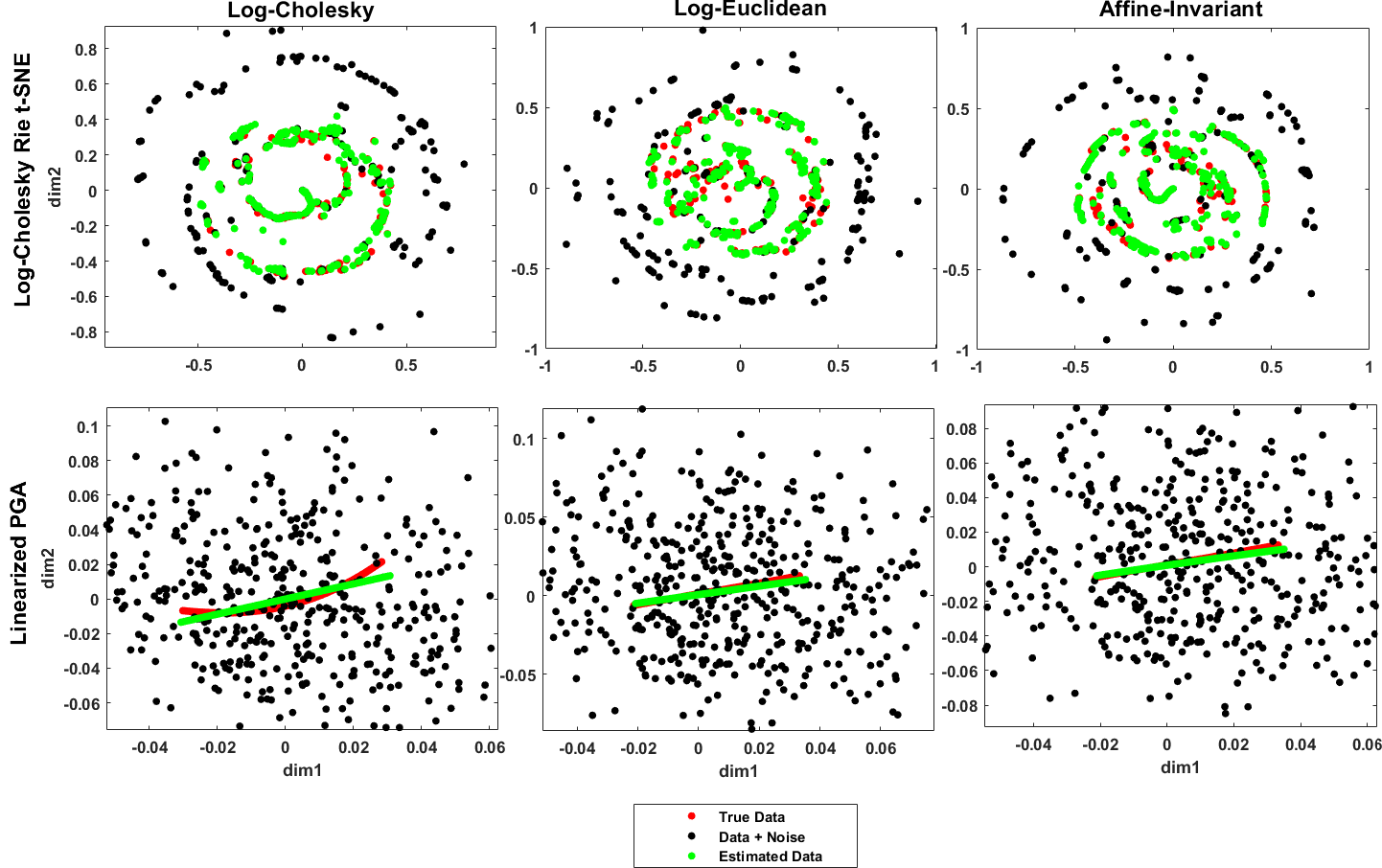}
	\caption{\small Visualization using Log-Cholesky Rie t-SNE and Linearized PGA for one realization of a Monte Carlo simulation SPD(18) with a scalar covariate. Log-Cholesky Rie t-SNE effectively captures nonlinearity and reveals intriguing spiral patterns even in noise, while Linearized PGA provides similar qualitative behavior. The TRUE DATA (red data) is close to the ESTIMATED DATA (green data) for all regression methods, indicating effective handling of noisy data}
	\label{fig:Rie_SNE}
\end{figure}

Visual exploration is often employed to gain insight into estimators' performance. However, the Ellipsoidal representation is no longer valid for high-dimensional SPD data. As a result, dimensionality reduction techniques, such as Log-Cholesky Rie t-SNE and Linearized PGA, are utilized. Our team has developed a novel visualization technique called Log-Cholesky Rie t-SNE, which involves performing Rie SNE on an SPD manifold equipped with the Log-Cholesky metric. This approach enables rapid computation of several features in the Rie-SNE pipeline.

Figure (\ref{fig:Rie_SNE}) depicts the results of performing Log-Cholesky Rie t-SNE (first row) and Linearized PGA (second row) for a realization in the Monte Carlo simulation on the manifold  SPD(18)   with a scalar covariate. The first row showcases the outcomes of Log-Cholesky Rie t-SNE for various regression techniques examined during the simulations. We observe that, regardless of the regression method, the TRUE DATA (red data) is near the ESTIMATED DATA (green data) and is further from the DATA+NOISE (black data). This implies that all regression methods effectively handle noisy data. When Linearized PGA is performed, the qualitative behavior is also similar. However, Log-Cholesky Rie t-SNE can handle the data's nonlinearity, even in the DATA+NOISE scenario, and captures intriguing spiral patterns.

\section{Discussion and Conclusion }

Our paper presents both theoretical and practical results. The theoretical results include general-purpose formulas for computing or expressing the multivariate Intrinsic Local Polynomial Regression (ILPR) estimator on the most convenient Isometric Riemannian Manifold (IRM). We define ILPR more formally than previous work, which allows for incorporating multivariate covariates and managing isometries on estimation problems. We establish a crucial theorem that relates the multivariate ILPR estimator on IRMs, making it easier to express the ILPR problem on a Riemannian Manifold with a more straightforward geometric structure.

Our research has resulted in a relevant theorem establishing a closed analytical expression for the multivariate ILPR estimator on Riemannian manifolds equipped with Euclidean Pullback Metrics (EPM). These metrics are obtained by pulling back the Euclidean metric from one manifold to another and covering an infinite family of Riemannian metrics. The closed expression of the multivariate ILPR estimator for EPM shares several similarities with the Local Polynomial Regression (LPR) estimator on Euclidean space \cite{ruppert1994multivariate,fan2018local,wand1994kernel}. For example, the expression of the intrinsic local constant estimator resembles the local constant estimator in Euclidean space, also known as the Nadaraya-Watson estimator \cite{fan2018local,wand1994kernel}.

Based on these similarities with classical results of non-parametric estimation, we established a formula for the asymptotic bias in the case of the induced estimator (Subsection \ref{relationsec}), as well as the consistency of the estimator under regularity conditions (Appendix \ref{regular}). These results are essential because many significant Riemannian metrics are, in fact, EPM. For instance, we prove that the Log-Cholesky metric \cite{lin2019riemannian} is an EPM along with the Log-Euclidean, Power-Euclidean, and Cholesky on the SPD Manifold, meaning that our results under regularity conditions imply the immediate consistency of the multivariate intrinsic local linear estimator for each of this metrics. 

We also wrote a pseudocode algorithm that implements the multivariate ILPR estimator on Riemannian manifolds equipped with EPM that summarizes the proposed method and could be used as a draft for further implementations. This algorithm and others that make simulations possible are available in MATLAB on the GitHub page. To demonstrate applications of the methodology, we conducted Monte Carlo simulations on the SPD manifold equipped with different metrics. We compared the performance of the multivariate ILPR for Log-Cholesky and Log-Euclidean Riemannian metrics with the well-established extrinsic LPR with the Affine-Invariant metric in the linear case \cite{barachant2010riemannian,li2022harmonized,lin2019extrinsic,sabbagh2020predictive}.

Simulation of DTI simulated data on the SPD(3) Manifold suggests that all regression methods considered are capable of handling noisy data. For instance, we visually appreciated the similarity in the Ellipsoidal Representation in Figure (\ref{fig:trade_off1}). On the other hand, assessing the performance by the Monte Carlo simulation on the SPD(3) manifold for a scalar covariate resulted that for DTI type data, the regression method with the best trade-off between error (RMSE) and time (CPU-Time)  was the intrinsic local linear regression with Log-Euclidean metric. This is an expected results from the literature \cite{arsigny2006log}. The method with the least RMSE score was the Affine-Invariant regression approach at the expense of a higher computational time. 

Subsequently, we examined the scalability of our findings to higher manifold and covariate dimensions. We formulated a mathematical model that generates a simulated high-dimensional regression problem considering the manifold covariate dimension to achieve this. We executed a Monte Carlo simulation for each manifold and covariate dimension within the specified range and calculated the RMSE and CPU-Time at the end of each iteration (Subsection \ref{simulations}). The results in Figure (\ref{fig:trade_off}) demonstrate that the earlier findings with DTI simulated data do not extend to higher manifold and covariate dimensions. In contrast, the Log-Cholesky metric yields a superior time-error trade-off for these more general scenarios. Finally, using Log-Cholesky Rie t-SNE and Linearized PGA, we show that as Ellipsoidal Representation for DTI simulated data, all the regression methods can handle noise data in the SPD manifold for higher dimension, in particular in SPD(18). 

A previous study, such as the one by Yuan \cite{yuan2012local}, examined the ILPR estimator's characteristics on the SPD manifold, but only for the Log-Euclidean and Affine-Invariant metrics, with a focus on univariate covariates. However, the method lacks a strict formal definition for the ILPR in the multivariate case and relies on parallel transport to the identity. In the univariate case, the study in \cite{yuan2012local} only obtained a closed analytical expression of the ILPR estimator for the Log-Euclidean metric. Although they conducted simulations and demonstrated that the ILPR estimator for the Log-Euclidean metric outperformed the ILPR for the Affine-Invariant metric, they also derived an asymptotic bias formula. They demonstrated the estimator's consistency and normality. Nevertheless, this research topic requires further investigation into the multivariate ILPR estimator's more general properties over an arbitrary Riemannian manifold, considering the presence of isometries.

Through our research, we have made significant strides toward understanding the theoretical and practical properties of the multivariate ILPR estimator. Our formal definition sets us apart from previous works on this topic, enabling us to tackle some of the problems inherent in the original formulation. Based on our discoveries, we could derive a closed analytical expression for the multivariate ILPR in the EPM family of metrics - an infinite set of metrics. Furthermore, we established the estimator's consistency in the local linear case and obtained a formula for its asymptotic bias for the EPM family. These results represent a significant advancement, expanding upon the previous findings and generalizing them to more complex settings.

The Monte Carlo simulations over the SPD manifold followed a similar approach in \cite{yuan2012local}. In the case of the DTI simulated data with $n=3$, the Log-Euclidean metric showed the best trade-off between time and error, consistent with previous findings. However, this performance did not scale when increasing the manifold and covariate dimension. In contrast, our new method for performing the multivariate ILPR estimator using the Log-Cholesky metric in the linear case demonstrated superior performance. This advancement provides a valuable addition to the available estimators for solving real data problems. Understanding how the regression problem scales with manifold and covariate dimensions is critical for practical applications, such as studying functional connectivity, EEG-cross spectra, or computer vision, where regression problems on SPD(n) often involve $n>>3$.

This study's theoretical and practical results have some limitations that must be acknowledged. The theorems established in this work rely on strong hypotheses, so their validity may be limited in some instances. Moreover, the closed analytical expression for the multivariate ILPR estimator only applies to the EPMs metric. The proof of the consistency of the estimator is only valid under restrictive regularity conditions for the local constant and linear cases. It would be interesting to investigate the validity of the estimator under more relaxed assumptions and to extend the proof of consistency to higher orders of regression (k > 1). Additionally, normality is a crucial statistical property of any estimator, and it would be valuable to derive and prove this asymptotic property for the multivariate ILPR estimator. For EPM, due to the resemblance of the closed analytical expression of the estimator to LPR in Euclidean space, it is possible to generalize other more advanced regression techniques using the idea of binning with FFT \cite{wang2022fkreg}.

Regarding the SPD manifold, we derived proof for a limited set of Riemannian metrics to be EPMs. However, many other metrics than the ones considered in Lemma (\ref{lemma1}) could also be EPMs \cite{dryden2009non,harandi2014manifold}. Not all Riemannian metrics can be EPMs, but we can combine these results with classical results in Riemannian Geometry, such as the Nash Embedding Theorem \cite{chen2004can}, to study the general properties of an arbitrary Riemannian metric.
The simulation study only tested the performance of the multivariate ILPR estimator for a limited set of metrics concerning extrinsic Local Polynomial Regression using the Affine-Invariant metric. The model used may not accurately reflect real-world scenarios, so a more general study is necessary using real data or a different model to generate true data. A more detailed study of how the regression method performs with increasing manifold dimensions is necessary. For example, typical regression problems on functional connectivity matrices require regression on the SPD$(n)$ with $n >> 18$.

The simulations were carried out with a fixed noise level ($\sigma^2 = 0.5$) to enable comparison with results reported in \cite{yuan2012local}, but it should be noted that the choice of noise level can impact the results. 
The simulation study was limited to one noise model Eq. (\ref{noisemodel}), which may only capture the noise characteristics in some applications. To ensure robust results, it is essential to investigate the sensitivity of the estimator to different Riemannian noise models at different noise levels, including specialized models such as the Riccian Noise Model in DTI \cite{schwartzman2006riemannian,Zhu2007b,yuan2012local}.
Additionally, the Least One Out Cross-Validation method used for bandwidth selection may need to be fully optimized and could be sensitive to outliers \cite{fan2018local,wand1994kernel}. Therefore, a robust Cross-Validation method may be necessary for large datasets. It would also be beneficial to explore computationally efficient bandwidth selection methods, such as k-Fold Cross-Validation \cite{fan2018local}, Generalized Cross Validation \cite{andrews1991asymptotic,li1986asymptotic}, or Akaike Information Criteria \cite{hurvich1998smoothing}. Finally, we presented Log-Cholesky Rie t-SNE, focusing only on its high dimensional SPD data application. Our presentation was simple and limited, and it is necessary to explore the method's performance on more complex tasks \cite{bergsson2022riesne}.

In the study conducted, the multivariate ILPR estimator, a statistical method utilized for regression analysis on symmetric positive definite (SPD) matrices, was investigated by researchers in more general settings. In specific cases, the Log-Cholesky metric demonstrated superior performance compared to other widely used metrics. Furthermore, the study emphasized the importance of comprehending how the regression problem scales with manifold and covariate dimensions, as it has practical implications in fields like computer vision and neuroscience. However, the study had limitations, such as relying on strong assumptions and a restricted set of metrics tested in simulations. In summary, the study provides valuable insights into utilizing the multivariate ILPR estimator in examining SPD matrices, but additional research is required to grasp its potential and limitations fully.

\begin{Backmatter}

\paragraph{Acknowledgments}
We are grateful for the technical assistance of Maria Luisa Bringas Vega.

\paragraph{Funding Statement}
This work was funded by the Chengdu Science and Technology Bureau Program under Grant 2022-GH02-00042-HZ. 

\paragraph{Competing Interests}
No competing interests

\paragraph{Data Availability Statement}
\url{https://github.com/ronald1129/Multivariate-Intrinsic-Local-Polynomial-Regression-on-Isometric-Riemannian-Manifolds.git}

\begin{appendix}
	
	\section{Definition of partial derivative respect to a vector}
	\begin{definition}[\bf Partial derivative of a matrix with respect to a vector]\label{deriv}
		The first order partial derivatives with respect to vector $X\in\mathbb{R}^p$ of scalar $\sigma(X)$, vector $s(X)\in \operatorname{M}_{n,1}(\mathbb{R})$ and matrix $S(X)\in \operatorname{M}_{n,m}(\mathbb{R})$ are respectively defined
		\begin{align}
			\begin{split}
				\partial_{X} \sigma(X)&=\begin{bmatrix}
					\frac{\partial \sigma(X)}{\partial x_1} &\frac{\partial \sigma(X)}{\partial x_2} &\cdots & \frac{\partial \sigma(X)}{\partial x_p}
				\end{bmatrix}\in \operatorname{M}_{1,p}(\mathbb{R}),
				\\	\partial_{X} s(X) &=\begin{bmatrix}
					\partial_{X} \operatorname{row}_1(s(X))\\ \partial_{X} \operatorname{row}_2(s(X))\\ \cdots \\ \partial_{X}\operatorname{row}_n(s(X)))
				\end{bmatrix} \in \operatorname{M}_{n,p}(\mathbb{R}),
				\\ \partial_{X} S(X)&= \begin{bmatrix}
					\partial_{X}\operatorname{col}_1(S(X)) & \partial_{X}\operatorname{col}_2(S(X)) & \cdots & \partial_{X}\operatorname{col}_m(S(X))
				\end{bmatrix} \in \operatorname{M}_{n,mp}(\mathbb{R}).
			\end{split}
		\end{align}
		
	\end{definition}
	\section{Proof of Lem. (\ref{lemma1})} \label{LogCholEPM}	
	\begin{proof}
		From Table  (\ref{Tab5}), it is evident that Lemma (\ref{lemma1}) is true for the  Riemannian Metric, Log-Euclidean, Power-Euclidean, Frobenius, Cholesky. Therefore only remains to prove that Log-Cholesky is also an EPM by $f(Y) = \lfloor\mathcal{L}(Y)\rfloor+\log\mathbb{D}(\mathcal{L}(Y))$. For that, we must prove that the following identity is true
		\begin{align}\label{intentytoprove}
			\begin{split}
				g^{LC}_Y(V,V) &= ||d_Yf(V)||_2^2,
				\\f(Y)& = \lfloor\mathcal{L}(Y)\rfloor+\log\mathbb{D}(\mathcal{L}(Y)).
			\end{split}
		\end{align}
		where $Y\in\operatorname{SPD}_n$, $V\in{T}_Y\operatorname{SPD}_n$and $g^{LC}$ is the Log-Cholesky metric defined in Table  (\ref{Tab5}). To prove this identity, previous results from the literature  are needed \cite{lin2019riemannian}
		\begin{align}\label{factL}
			{d}_Y\mathcal{L}(V) &= \mathcal{L}(Y)(\mathcal{L}(Y)^{-1}V\mathcal{L}(Y)^{-T})_{1/2},
			\\ \label{facttaylor}\mathcal{L}(Y+\epsilon V)& = \mathcal{L}_Y(V) +\epsilon d_Y\mathcal{L}(V)+ \text{O}(\epsilon^2),
			\\\label{fact1/2}(Y)_{1/2} & = \lfloor Y \rfloor + \mathbb{D}(Y)/2,
			\\\label{factD+}\mathbb{D}(L_1+L_2) &= \mathbb{D}(L_1)+\mathbb{D}(L_2), 
			\\\label{factDx} \mathbb{D}(L_1 L_2) &= \mathbb{D}(L_1)\mathbb{D}(L_2),
			\\\label{factcomm}\mathbb{D}(D_1)\mathbb{D}(D_2) & = \mathbb{D}(D_2)\mathbb{D}(D_1)
			\\\label{factlog} \log(\mathbb{D}(D_1)\mathbb{D}(D_2)) &= \log\mathbb{D}(D_1)+\log\mathbb{D}(D_2),
			\\\label{factL+} \lfloor L_1 +L_2 \rfloor & = \lfloor L_1 \rfloor + \lfloor L_2 \rfloor,
		\end{align}
		where  as before $Y\in \text{SPD}_n$, $V\in{T}_Y\text{SPD}_n$ and $D_{1},D_2\in \text{Diag}_n$,  $L_1,L_2\in\text{LT}_n$.
		From this identities the differential ${d}_Yf(V)$  can be computed
		\begin{align}\label{startdifferential}
			\begin{split}
				{d}_Yf(V) &= \lim_{\epsilon\to0}\epsilon^{-1}\left(\lfloor\mathcal{L}(Y+\epsilon V)\rfloor+\log\mathbb{D}(\mathcal{L}(Y+\epsilon V))- \lfloor\mathcal{L}(Y)\rfloor-\log\mathbb{D}(\mathcal{L}(Y))\right) 
				\\&= \lim_{\epsilon\to 0}\epsilon^{-1}\left(\lfloor\mathcal{L}(Y+\epsilon V)\rfloor - \lfloor\mathcal{L}(Y)\rfloor+\log\mathbb{D}(\mathcal{L}(Y+\epsilon V)) - \log\mathbb{D}(\mathcal{L}(Y))\right).
			\end{split}
		\end{align}
		To continue a first order Taylor expansion respect $\epsilon$ is needed for the increments $\lfloor\mathcal{L}(Y+\epsilon V)\rfloor - \lfloor\mathcal{L}(Y)\rfloor$  and $\log\mathbb{D}(\mathcal{L}(Y+\epsilon V)) - \log\mathbb{D}(\mathcal{L}(Y))$. 
		\begin{align}\label{dl}
			\begin{split}
				\lfloor\mathcal{L}(Y+\epsilon V)\rfloor - \lfloor\mathcal{L}(Y)\rfloor & =^{\text{by \ref{factL+}}} \lfloor\mathcal{L}(Y+\epsilon V)-\mathcal{L}(Y)\rfloor
				\\ & \approx^{\text{by \ref{facttaylor}}} \lfloor\mathcal{L}(Y)+\epsilon d_Y\mathcal{L}(V)  - \mathcal{L}(Y)\rfloor
				\\&  =^{\text{by \ref{factL}}}  \lfloor\mathcal{L}(Y)+\epsilon \mathcal{L}(Y)(\mathcal{L}(Y)^{-1}V\mathcal{L}(Y)^{-T})_{1/2}  - \mathcal{L}(Y)\rfloor
				\\& =  \epsilon\lfloor \mathcal{L}(Y)(\mathcal{L}(Y)^{-1}V\mathcal{L}(Y)^{-T})_{1/2}\rfloor.
			\end{split}
		\end{align}
		On the other hand the first order Taylor expansion of  $\log\mathbb{D}(\mathcal{L}(Y+\epsilon V)) - \log\mathbb{D}(\mathcal{L}(Y))$ is  
		\begin{eqnarray}\label{dlog}
			\begin{split}
				&\log\mathbb{D}(\mathcal{L}(Y+\epsilon V)) - \log\mathbb{D}(\mathcal{L}(Y)) =^{\text{by \ref{factlog}}}  \log\left(\mathbb{D}(\mathcal{L}(Y+\epsilon V))\mathbb{D}(\mathcal{L}(Y))^{-1}\right) 
				\\ &= ^{\text{by \ref{facttaylor}}}    \log\left(\mathbb{D}\left(\mathcal{L}(Y)+\epsilon d_Y\mathcal{L}(V)+\text{O}(\epsilon^2)\right)\mathbb{D}(\mathcal{L}(Y))^{-1}\right) 
				\\ & = ^{\text{ by \ref{factL}}}   \log\left(\mathbb{D}\left(\mathcal{L}(Y)+\epsilon  \mathcal{L}(Y)(\mathcal{L}(Y)^{-1}V\mathcal{L}(Y)^{-T})_{1/2} +\text{O}(\epsilon^2)\right)\mathbb{D}(\mathcal{L}(Y))^{-1}\right) 
				\\& = ^{\text{ by \ref{factD+}}}  \log\left(\left(\mathbb{D}(\mathcal{L}(Y))+\epsilon  \mathbb{D}(\mathcal{L}(Y)(\mathcal{L}(Y)^{-1}V\mathcal{L}(Y)^{-T})_{1/2}) +\text{O}(\epsilon^2)\right)\mathbb{D}(\mathcal{L}(Y))^{-1}\right)
				\\& = ^{\text{ by \ref{factDx}}} \log\left(\left(\mathbb{D}(\mathcal{L}(Y))+\epsilon  \mathbb{D}(\mathcal{L}(Y))\mathbb{D}((\mathcal{L}(Y)^{-1}V\mathcal{L}(Y)^{-T})_{1/2}) +\text{O}(\epsilon^2)\right)\mathbb{D}(\mathcal{L}(Y))^{-1}\right)
				\\& =  ^{\text{ by \ref{factcomm}}} \log\left(I_n+\epsilon \mathbb{D}((\mathcal{L}(Y)^{-1}V\mathcal{L}(Y)^{-T})_{1/2}) +\text{O}(\epsilon^2)\right)
				\\& \approx \epsilon \mathbb{D}((\mathcal{L}(Y)^{-1}V\mathcal{L}(Y)^{-T})_{1/2}).
			\end{split}
		\end{eqnarray}
		Combining {Eqs}.  (\ref{dl},\ref{dlog}) in {Eq}. (\ref{startdifferential}) result
		\begin{align}
			\begin{split}
				{d}_Yf(V) &=   \lim_{\epsilon\to0}\epsilon^{-1}\left(\lfloor\mathcal{L}(Y+\epsilon V) - \mathcal{L}(Y)\rfloor+\log\mathbb{D}(\mathcal{L}(Y+\epsilon V)) - \log\mathbb{D}(\mathcal{L}(Y))\right)
				\\& =  \lim_{\epsilon\to 0}\epsilon^{-1}\left(\epsilon\lfloor \mathcal{L}(Y)(\mathcal{L}(Y)^{-1}V\mathcal{L}(Y)^{-T})_{1/2}\rfloor +\epsilon \mathbb{D}((\mathcal{L}(Y)^{-1}V\mathcal{L}(Y)^{-T})_{1/2})\right)
				\\& =\lfloor \mathcal{L}(Y)(\mathcal{L}(Y)^{-1}V\mathcal{L}(Y)^{-T})_{1/2}\rfloor +\mathbb{D}((\mathcal{L}(Y)^{-1}V\mathcal{L}(Y)^{-T})_{1/2}).
			\end{split}
		\end{align}
		From this we can finally prove the identity { Eq.} (\ref{intentytoprove})
		\begin{align}
			\begin{split}
				||{d}_Yf(V)||_2^2 &= ||\lfloor \mathcal{L}(Y)(\mathcal{L}(Y)^{-1}V\mathcal{L}(Y))^{-T})_{1/2}\rfloor + \mathbb{D}((\mathcal{L}(Y)^{-1}V\mathcal{L}(Y)^{-T})_{1/2})||_2^2
				\\& = ||\lfloor \mathcal{L}(Y)(\mathcal{L}(Y)^{-1}V\mathcal{L}(Y)^{-T})_{1/2}\rfloor||_2^2 +|| \mathbb{D}((\mathcal{L}(Y)^{-1}V\mathcal{L}(Y)^{-T})_{1/2})||_2^2
				\\& = g^{LC}_Y(V,V).
			\end{split}
		\end{align}
	\end{proof}
	\section{Proof Lem. (\ref{Lemmaiso})} \label{Lemmaproof}
	\begin{proof}
		If $L: \mathcal{A}\subset \operatorname{M}_n(\mathbb{R}) \to \operatorname{M}_n(\mathbb{R})$  is a linear invertible map, then by the Matrix Representation of Linear Transformations  there is a matrix $M\in \operatorname{GL}_{n^2}(\mathbb{R})$ such that 
		\begin{align}
			\operatorname{vec}(L(Z)) = M \operatorname{vec}(Z), \,\,\, \forall Z\in \mathcal{A}. 
		\end{align} 
		On the other hand, by the Kronecker Product Decomposition \cite{liu2012new,mezzarkronecker}, there are two invertible matrices $A$ and $B \in \operatorname{GL}_n(\mathbb{R})$ such that 
		\begin{align}
			M = B^T \otimes A,
		\end{align}
		therefore
		\begin{align}
			\begin{split}
				L(Z)& =\operatorname{vec}^{-1}\left(\operatorname{vec}(L(Z))\right)
				\\&=\operatorname{vec}^{-1}\left(M\operatorname{vec}(Z)\right)
				\\& = \operatorname{vec}^{-1}\left((B^T\otimes A)\operatorname{vec}(Z)\right)
				\\& = \operatorname{vec}^{-1}\left(\operatorname{vec}(AZB)\right)
				\\& = AZB.
			\end{split}
		\end{align}
	\end{proof}
	\section{Proof of Thm. (\ref{MILPR})}\label{appendixA}

	\begin{proof}
		Let us denote the exponential, logarithm map, the parallel transport, and the distance induced from the metric $G$ on the Riemannian  Matrix Manifold $(\mathcal{M}_n, g^f)$  by $\operatorname{exp}^f_{Y }(\cdot)$, $\operatorname{log}^f_{Y }(\cdot)$, $\pi^f_{Y_1\to Y_2 }$  and $\operatorname{dist}_{g^f}$. Let us also denote equivalent operation induced from the metric $g$ on the Riemannian Matrix Manifold $(f(\mathcal{M})_n,g)$ by $\exp_{f(Y)}(\cdot)$, $\log_{f(Y)}(\cdot)$, $\pi_{f(Y_1)\to f(Y_2)}$  and $\operatorname{dist}_g$. Because $f$ is an isometry by hypothesis, $g^f$ is a pullback metric of $g$ induced by  $f$. Then  the following relations holds,  {Table}  (\ref{Tab1})
		\begin{align} \label{eqt1}
			\operatorname{exp}^f_{Y}(V)&=f^{-1}\left(\exp_{f(Y)}\left({d}_{Y}f(V)\right)\right),
			\\ \label{eqt2}\operatorname{log}^f_{Y_1}(Y_2)&={d}_{f(Y_1)}f^{-1}\left(\log_{f(Y_1)}f(Y_2)\right),
			\\ \label{eqt3}\operatorname{dist}_{g^f}(Y_1,Y_2)&=\operatorname{dist}_g(f(Y_1),f(Y_2)),
			\\ \pi^f_{Y_1\rightarrow Y_2}V&={d}_{f(Y_2)}f^{-1}\left(\pi_{f(Y_1)\rightarrow f(Y_2)}({d}_{Y_1}f(V))\right)\label{eqt4},
		\end{align}   
		where $Y,Y_1, Y_2 \in\mathcal{M}_n$, $f(Y_1),f(Y_1), f(Y_2)\in f(\mathcal{M}_n)$ and $V\in {T}_Y\mathcal{M}_n$.
		To understand the development of the proof, we must bear in mind that the applications 
		\begin{align}\label{linearpopertie}
			\begin{split}
				{d}_Yf:& T_S\mathcal{M}_n\rightarrow T_{f(S)}f(\mathcal{M}_n),
				\\	d_{f(Y)}f^{-1}:& T_{f(Y)}f(\mathcal{M}_n)\rightarrow T_{Y}\mathcal{M}_n,
				\\\pi^f_{Y_1\to Y_2}:& T_{Y_1}\mathcal{M}_n\rightarrow{T}_{Y_2}\mathcal{M}_n,
				\\\pi_{f(Y_1)\to f(Y_2)}:& T_{f(Y_1)}f(\mathcal{M}_n)\rightarrow {T}_{f(Y_2)}f(\mathcal{M}_n),
			\end{split}	
		\end{align}
		are all invertible linear maps that satisfy the properties
		\begin{align} \label{inversionpopertie}
			\begin{split}
				\left(d_Yf\right)^{-1}&= d_{f(Y)}f^{-1},
				\\(\pi^f_{Y_1\to Y_2})^{-1} &= \pi^f_{Y_1\to Y_2},
				\\(\pi_{f(Y_1)\to f(Y_2)})^{-1}& = \pi_{f(Y_2)\to f(Y_1)}.
			\end{split}
		\end{align}
		By hypothesis $b)$, the ILE  $\widehat{\alpha}(x;k,h)$ exists and is defined as the solution of the following optimization problem Definition (\ref{def3})
		
		\begin{align}\label{opt1}
			\begin{split}
				\widehat{\alpha}(x;k,h)& = \operatorname{arg} \max_{\alpha} G_1(\alpha,x,\omega,k,h,\mathbf{\Sigma},\mathcal{M}_n,g^f),
				\\G_1(\alpha,x,\omega,k,h,\mathbf{\Sigma},\mathcal{M}_n,G)&=\sum _{i=1}^{N}K_{i,h}(x)\operatorname{dist}_{g^f}^2\left(m^{[k]}_{\alpha,x}(X_i),Y_i\right),
			\end{split}
		\end{align}
		where $K_{i,h}(x)=K_h(X_i-x )$ for some kernel function $K_{h}(\cdot)$   with bandwidth $h$, $\operatorname{dist}_{g^f}$ is the Riemannian distance associated  to $g^f$,    $\alpha=(\alpha_0,\alpha_1,\alpha_2,\cdots,\alpha_k)\in \mathcal{M}_n\times \text{M}_{n,np}(\mathbb{R})  \times \cdots \times \text{M}_{n,np^k}(\mathbb{R}) $ and $m_{\alpha,x}^{[k]}(X)$ is the Intrinsic Polynomial described by 
		\begin{align}\label{poly1}
			\begin{split}
				m_{\alpha,x}^{[k]}(X)&=^{\text{by \ref{def3}}} \operatorname{exp}^f_{\alpha_0}\left(\pi^f_{\omega\to\alpha_0}t_{\alpha,x}^{[k]}(X)\right),
				\\t_{\alpha,x}^{[k]}(X) & = \sum_{j=1}^{k} \alpha_j (I_n\otimes(X - x)^{\otimes j}).
			\end{split}
		\end{align}
		By the hypothesis $b)$, the ILE $\widehat{\beta}$ exists and is defined as the solution of the following optimization problem Definition (\ref{def3})
		\begin{align}\label{opt2}
			\begin{split}
				\widehat{\beta}& = \operatorname{arg} \max_{\beta} G_2(\beta,x,f(\omega),k,h,f(\mathbf{\Sigma}),f(\mathcal{M}_n),g),
				\\G_2(\beta,x,f(\omega),k,h,f(\mathbf{\Sigma}),f(\mathcal{M}_n),g)&=\sum _{i=1}^{N}K_{i,h}(x)\operatorname{dist}_g^2\left((fm)^{[k]}_{\beta,x}(X_i),f(Y_i)\right),
			\end{split}
		\end{align}
		whereas before $K_{i,h}(x)=K_h(X_i-x )$ for some kernel function $K_{h}(\cdot)$   with bandwidth $h$, $d$ is the Riemannian distance associated  to $g$, $\beta=(\beta_0,\beta_1,\beta_2,\cdots,\beta_k)\in f(\mathcal{M}_n)\times \text{M}_{n,np}(\mathbb{R})  \times \cdots \times \text{M}_{n,np^k}(\mathbb{R}) $ and  $(fm)_{\beta,x}^{[k]}(X)$ is the Intrinsic Polynomial described by
		\begin{align}\label{poly2}
			\begin{split}
				(fm)_{\beta,x}^{[k]}(X)&=  ^{\text{by \ref{def3}}} \operatorname{exp}_{\beta_0}\left(\pi_{f(\omega)\to\beta_0} t_{\beta,x}^{[k]}(X)\right),
				\\t_{\beta,x}^{[k]}(X) & = \sum_{j=1}^{k} \beta_j (I_n\otimes(X - x)^{\otimes j}).
			\end{split}
		\end{align}

		To obtain the thesis of the Theorem  {Eq}. (\ref{polypoly}), we must find a way to relate the problems {Eqs}.  (\ref{opt1}, \ref{opt2}), this means that we must find a change of variable $\alpha = \phi (\beta)$, in such a way that $\phi:f(\mathcal{M}_n)\times \text{M}_{n,np}(\mathbb{R})  \times \cdots \times \text{M}_{n,np^k}(\mathbb{R}) \to \mathcal{M}\times \text{M}_{n,np}(\mathbb{R})  \times \cdots \times \text{M}_{n,np^k}(\mathbb{R})$ is a diffeomorphism and with this we guarantee that $\widehat{\alpha} = \phi (\widehat{\beta})$. To construct such diffeomorphism, first, we have noted that  {Eqs}.  (\ref{eqt3}, \ref{eqt1}, \ref{poly1}) implies that we can manipulate the optimization function $G_1(\alpha,x,\omega,k,h,\mathbf{\Sigma},\mathcal{M}_n,g^f)$ in {Eq}. (\ref{opt1}) in the following manner 
		\begin{align}\label{notice1}
			\begin{split}
				G_1(\alpha,x,\omega,k,h,\mathbf{\Sigma},\mathcal{M}_n,g^f)&=\sum _{i=1}^{N}K_{i,h}(x)\operatorname{dist}_{g^f}^2\left(m^{[k]}_{\alpha,x}(X_i),Y_i\right)
				\\&=^{\text{by \ref{eqt3}}}\sum _{i=1}^{N}K_{i,h}(x)\operatorname{dist}_g^2\left(f\left(m^{[k]}_{\alpha,x}(X_i)\right),f(Y_i)\right)
				\\&=^{\text{by \ref{eqt1},\ref{poly1}}}\sum _{i=1}^{N}K_{i,h}(x)\operatorname{dist}_g^2\left(\exp_{f(\alpha_0)}\left(d_{\alpha_0} f \pi^f_{\omega\to\alpha_0 }t_{\alpha,x}^{[k]}(X)\right),f(Y_i)\right), 
			\end{split}
		\end{align}
		now if we look closely at the last expression, we can notice that this expression reassembles the optimization function $G_2(\beta,x,f(\omega),k,h,f(\mathbf{\Sigma}),f(\mathcal{M}_n),g)$ in {Eq}. (\ref{opt2}), however, it does not match completely due to the first term inside the distance involving exponential map, etc. On the other hand, we can make these expressions match by choosing, if possible, a change of variable $\alpha=\phi(\beta)$ in such a way that the problematic term in {Eq}. (\ref{notice1}) is precisely the Intrinsic Polynomial $(fm)_{\beta,x}^{[k]}(X)$ in {Eq}. (\ref{poly2}), that is
		\begin{align}
			\label{cerotransf}
			\alpha_0 &=f^{-1}(\beta_0),
			\\\label{changevariablecondition}\exp_{f(\alpha_0)}\left({d}_{\alpha_0} f \pi^f_{\omega\to\alpha_0 }t_{\alpha,x}^{[k]}(X)\right)&= \operatorname{exp}_{\beta_0}\left(\pi_{f(\omega)\to\beta_0} t_{\beta,x}^{[k]}(X)\right), \,\,\, \forall X\in\mathbb{R}^p.
		\end{align}
		With this, we already know how the first components of $\alpha$ and $\beta$ are transformed by $\phi$. All remains to establish how the rest of the components $\alpha_j$ and $\beta_j$ for $j=1,2,\cdots,k$ are related. Using {Eq}. (\ref{cerotransf}) in {Eq}. (\ref{changevariablecondition}) we can eliminate the exponential operators and be left with the arguments of {Eq}. (\ref{changevariablecondition}), that is 
		\begin{align}\label{changevariablecondition2}
			d_{\alpha_0} f \pi^f_{\omega\to\alpha_0 }t_{\alpha,x}^{[k]}(X)= \pi_{f(\omega)\to\beta_0} t_{\beta,x}^{[k]}(X), \,\,\, \forall X\in\mathbb{R}^p,
		\end{align}
		this means that the change of variables we are looking for $\phi$ is such that it satisfies {Eqs}.  (\ref{changevariablecondition2}, \ref{cerotransf}).  Now taking into account the properties {Eqs}.  (\ref{eqt4}, \ref{cerotransf}, \ref{inversionpopertie}) we have 
		\begin{align}\label{inversion1}
			\begin{split}
				\left({d}_{\alpha_0} f \pi^f_{\omega\to\alpha_0 }\right)^{-1} \pi_{f(\omega)\to\beta_0}&=^{\text{by \ref{eqt4}}}	\left(\pi_{f(\omega)\to f(\alpha_0)} d_{\omega}f\right)^{-1} \pi_{f(\omega)\to\beta_0}
				\\ &=^{\text{by \ref{cerotransf},\ref{inversionpopertie}}}\left( d_{f(\omega)}f^{-1}\pi_{\beta_0\to f(\omega)}\right)\pi_{f(\omega)\to\beta_0}
				\\&= d_{f(\omega)}f^{-1}\left(\pi_{\beta_0\to f(\omega)}\pi_{f(\omega)\to\beta_0}\right)
				\\ &= d_{f(\omega)}f^{-1}.
			\end{split}
		\end{align}
		Using  {Eq}. (\ref{inversion1}) we can rewrite {Eq}. (\ref{changevariablecondition2}) as follows
		\begin{align}\label{changeofvariable3}
			\begin{split}
				t_{\alpha,x}^{[k]}(X)&= d_{f(\omega)}f^{-1} t_{\beta,x}^{[k]}(X)
				\\&=L\left(t_{\beta,x}^{[k]}(X)\right), \,\,\, \forall X\in\mathbb{R}^p,
			\end{split}
		\end{align}
		where $L:{T}_{f(\omega)}f(\mathcal{M}_n)\to T_{\omega}\mathcal{M}_n$ is an invertible linear map  because is the composition of  invertible  linear maps
		\begin{align}
			L= d_{f(\omega)}f^{-1}.
		\end{align} 
		Since $(\mathcal{M}_n,g^f)$ and $(f(\mathcal{M}_n),g)$ are Riemannian Matrix Manifold, $f(\mathcal{M}_n)\subseteq \text{M}_n(\mathbb{R})$  and  the tangent spaces ${T}_{\omega}\mathcal{M}_n$ and ${T}_{f(\omega)}f(\mathcal{M}_n)$  are vector subspaces of $\text{M}_n(\mathbb{R})$, therefore $L$ is an invertible linear map between subspaces of $\text{M}_n(\mathbb{R })$.  This means that the hypothesis of the Lemma (\ref{Lemmaiso}) are satisfied, therefore  there are  two invertible matrices $A, B \in \text{GL}_n(\mathbb{R})$ such that 
		\begin{align}\label{lpropertie}
			L(Z)=AZB,
		\end{align}
		combining with the result {Eq}. (\ref{lpropertie}, \ref{changeofvariable3}) we get
		\begin{align}\label{change4}
			t_{\alpha,x}^{[k]}(X)=At_{\beta,x}^{[k]}(X)B, \,\,\, \forall X \in\mathbb{R}^p.
		\end{align}
		To find the relationship between the coefficients $\alpha_l$ and $\beta_l$, we only have to differentiate {Eq}. (\ref{change4}) taking into account {Eqs}.  (\ref{taylorcoef}, \ref{productderivate}) 
		\begin{align}\label{computo1}
			\begin{split}
				\alpha_l &=^{\text{by }\ref{taylorcoef}}\frac{1}{l!}{\partial}_{x}^lt_{\alpha,x}^{[k]}(X)
				\\&=^{\text{by \ref{change4}}}	\frac{1}{l!}{\partial}_{x}^l \left(At_{\beta,x}^{[k]}(X)B\right)
				\\&=^{\text{by \ref{productderivate}}}	A\left(\frac{1}{l!}{\partial}_{x}^l \left(\sum_{j=1}^{k}\beta_j(I_n\otimes(X-x)^{\otimes j})\right)\right)\left(B \otimes I_p^{\otimes l}\right)
				\\&=^{\text{by }\ref{taylorcoef}} A \beta_l \left(B \otimes I_p^{\otimes l}\right), \,\,\, \forall l=1,2,\cdots,  k. 
			\end{split}
		\end{align} 
		{ Eqs}. (\ref{computo1}, \ref{cerotransf}) implies that the map $\phi$  
		\begin{align}\label{difeo}
			\begin{split}
				\phi: f(\mathcal{M}_n)\times \text{M}_{n,np}(\mathbb{R})  \times \cdots \times \text{M}_{n,np^k}(\mathbb{R}) &\to   \mathcal{M}_n\times \text{M}_{n,np}(\mathbb{R})  \times \cdots \times \text{M}_{n,np^k}(\mathbb{R})
				\\  \beta=(\beta_0,\beta_1,\beta_2,\cdots,\beta_k)   &\to \alpha=(\alpha_0,\alpha_1,\alpha_2,\cdots,\alpha_k),  
				\\\alpha_j = \delta_{j0}f^{-1}(\beta_j)+& (1-\delta_{j0})A \beta_j \left(B \otimes I_p^{\otimes j}\right), \,\,\, \forall j=0,1,2,\cdots, k
			\end{split}
		\end{align}
		where $A$ and $B$ are the invertible matrices defined in {Eq}. (\ref{lpropertie}) and $\delta_{ij}$ is the Kronecker delta, is a diffeomorphism  such that 
		\begin{align}
			\begin{split}
				\widehat{\alpha}&=\phi(\widehat{\beta}),
				\\G_1(\phi(\widehat{\beta}),x,\omega,k,h,\boldsymbol{\Sigma},\mathcal{M}_n,g^f) &=  G_2(\widehat{\beta},x,f(\omega),k,h,f(\boldsymbol{\Sigma}),f(\mathcal{M}_n),g),
				\\\label{contruct}d_{\widehat{\alpha}_0} f \pi^f_{\omega\to \widehat{\alpha}_0 }t_{\widehat{\alpha},x}^{[k]}(X)&= \pi_{f(\omega)\to\widehat{\beta}_0} t_{\widehat{\beta},x}^{[k]}(X), \,\,\, \forall X\in\mathbb{R}^p.
			\end{split}
		\end{align}
		Therefore we conclude the main thesis of the Theorem  
		\begin{align}
			\widehat{\alpha}_j (x;k,h) = \delta_{j0}f^{-1}(\widehat{\beta}_j(x;k,h))+& (1-\delta_{j0})A \widehat{\beta}_j(x;k,h) \left(B \otimes I_p^{\otimes j}\right), \,\,\, \forall j=0,1,2,\cdots, k
		\end{align}
	\end{proof}
	
	\section{Proof Theorem  (\ref{teorema3})}\label{Proofteorem3}
	\begin{proof}
		To make the proof easier to understand, we will divide this proof into two parts in which each of the statements of the Theorem is demonstrated.
		
		\subsection{Expression of the ILP  {Eq}. (\ref{s1})}
		
		By hypothesis, the Riemannian Matrix  Manifold $(\mathcal{M}_n,g^f)$ and $(f(\mathcal{M}_n),g^E)$ are IRMs by isometry $f$, this mean by Definition (\ref{DeformedRiemannianManifold}) that  $g^f$  is the pullback metric of the Euclidean one. Then hypothesis of  Theorem (\ref{MILPR}) holds. Then invoking this Theorem  in particular  for $j=0$  in Eq. (\ref{polypoly}) implies that 
		\begin{align}\label{c1}
			\widehat{\alpha}_0(x;k,h) = f^{-1}(\widehat{\beta}_0(x;k,h)).
		\end{align}
		\subsection{Exact analytical expression of the ILE {Eq}. (\ref{s3})}
		By hypothesis $\widehat{\beta}(x;k,h)=\text{ILE}(x, f(\omega),k,h,f(\mathbf{\Sigma}),f(\mathcal{M}_n),g^E)$ exist, this means that $\widehat{\beta}$  is by Definition (\ref{def3}) the solution of the following optimization problem
		\begin{align}\label{c4}
			\begin{split}
				\widehat{\beta}(x;k,h)& = \operatorname{arg} \max_{\beta} G_2(\beta,x,f(\omega),k,h,f(\mathbf{\Sigma}),f(\mathcal{M}_n),g^E),
				\\G_2(\beta,x,f(\omega),k,h,f(\mathbf{\Sigma}),f(\mathcal{M}_n),g^E)&=\sum _{i=1}^{N}K_{i,h}(x)\operatorname{dist}_{g^E}^2\left((fm)^{[k]}_{\beta,x}(X_i),f(Y_i)\right),
			\end{split}
		\end{align}
		now because by hypothesis $\operatorname{dist}_{g^E}$ is the induced distance by the Frobenius norm  $g^E$ ($\operatorname{dist}_{g^E}^2(x,y)=||x-y||_2^2$). Now we can reduce the optimization problem {Eq}. (\ref{c4})  to the following one 
		\begin{align}\label{c6}
			\begin{split}
				\widehat{\beta}&=\operatorname{arg}\max_{\beta}\sum _{i=1}^{N}K_{i,h}(x)\operatorname{dist}_{g^E}^2\left((fm)^{[k]}_{\beta,x}(X_i),f(Y_i)\right)
				\\ &= \operatorname{arg}\max_{\beta}\sum _{i=1}^{N}K_{i,h}(x) || -f(Y_i)+\sum_{j=0}^{k}{\beta}_j X^j_{x,n}(X_i)||_2^2,
			\end{split}
		\end{align}
		where we denote $\widehat{\beta}=\widehat{\beta}(x;k,h)$ and $X_{x,n}^j(X_i) = I_n\otimes (X_i-x)^{\otimes j}$ for simplicity.  To continue, we only have to compute the gradient of the optimization function in {Eq}. (\ref{c6}), so for that lets denote $Z_i=-f(Y_i)+\sum_{j=0}^{k}{\beta}_j X^j_{x,n}(X_i)$, then 
		\begin{align}\label{c7}
			d_{\beta_l}Z_i=d(\beta_l X^{l}_{x,n} (X_i)),
		\end{align}
		using this fact, we compute the differential of the optimization function 
		\begin{eqnarray}
			\begin{split}
				d_{\beta_l}\left( \sum _{i=1}^{N}K_{i,h}(x) || -f(Y_i)+\sum_{j=0}^{k}{\beta}_j X^j_{x,n}(X_i)||_2^2 \right)&= d_{\beta_l}\left( \sum _{i=1}^{N}K_{i,h}(x) Z_i:Z_i\right)
				\\&=2\sum _{i=1}^{N}K_{i,h}(x) d_{\beta_l} (Z_i:Z_i)
				\\&=2\sum _{i=1}^{N}K_{i,h}(x)Z_i:d_{\beta_l}Z_i
				\\&=^{\text{by \ref{c7}}} 2 \sum _{i=1}^{N}K_{i,h}(x)Z_i:d(\beta_l X^{l}_{x,n} (X_i))
				\\&=2\sum _{i=1}^{N}K_{i,h}(x)(Z_i X^{l}_{x,n} (X_i)^T):d\beta_l
				\\&=2\left(\sum _{i=1}^{N}K_{i,h}(x)\left(-f(Y_i)+\sum_{j=0}^{k}{\beta}_j X^j_{x,n}(X_i)\right) X^{l}_{x,n} (X_i)^T\right):d\beta_l,
			\end{split}
		\end{eqnarray}
		therefore the gradient of the optimization function is 
		\begin{align}
			\frac{\partial}{\partial \beta _l} \left( \sum _{i=1}^{N}K_{i,h}(x) || -f(Y_i)+\sum_{j=0}^{k}{\beta}_j X^j_{x,n}(X_i)||_2^2 \right) = 2\sum _{i=1}^{N}K_{i,h}(x)\left(-f(Y_i)+\sum_{j=0}^{k}{\beta}_j X^j_{x,n}(X_i)\right) X^{l}_{x,n} (X_i)^T,
		\end{align}
		where $l=0,1,2,\cdots, k $. Finally, to find $\widehat{\beta}$, we only have to make the gradient equal to zero, with which we obtain the following system of $k+1$ matrix equations	 
		\begin{align}
			\sum _{i=1}^{N}K_{i,h}(x)f(Y_i)X^{l}_{x,n} (X_i)^T=\sum_{j=0}^{k}{\widehat{\beta}}_j \sum _{i=1}^{N}K_{i,h}(x)  X^j_{x,n}(X_i) X^{l}_{x,n} (X_i)^T, \,\,\,  l=0,1,2,\cdots,k.
		\end{align}
		or more explicit form
		\begin{align}\label{explicit}
			\sum _{i=1}^{N}K_{i,h}(x)f(Y_i)\left(I_n\otimes (X_i-x)^{\otimes l}\right)^T =\sum_{i=1 }^{N} \sum_{j=0}^{k}{\widehat{\beta}}_jK_{i,h}(x)  \left(I_n\otimes (X_i-x)^{\otimes j}\right) \left(I_n\otimes (X_i-x)^{\otimes l}\right)^T, 
		\end{align}
		where $l = 0,1,2,\cdots , k.$
		
		To prove the final statement of the Theorem, we need to solve the system of $k+1$ matrix equations in $\widehat{\beta}$ {Eqs}.  (\ref{explicit}). To archive this goal, we will be employing the following identity of linear algebra 
		\begin{align}\label{algeidenblock}
			\sum_{i =1}^{N}\lambda_i \text{colb}_i(A_{r\times m})\text{rowb}_i(B_{m\times s}) = A_{r\times m} \left(I_{m/N}\boxtimes \text{Diag}(\lambda_1,\lambda_2,\cdots,\lambda_N)\right)B_{m\times s}, 
		\end{align} 
		that is true for all block matrices $A$ and $B$  if $m/N$ is an integer and the number of elements in a block column of $A$ is equal to the number of elements in the correspondent block row of $B$. On the other hand from the Definition  of $\widehat{\beta}$, $f(\mathbf{Y})$ and $\mathbf{X}$ at the end of the Theorem  we conclude that: 
		\begin{align}
			\text{colb}_j(\widehat{\beta}) & = \widehat{\beta}_j \in \text{M}_{n,np^{j}}(\mathbb{R}), \,\,\,  j =1,2,\cdots, k, \label{columnb}
			\\ \text{colb}_i(f(\mathbf{Y})) & = f(Y_i) \in \text{M}_{n,n}(\mathbb{R}), \,\,\,  i =1,2,\cdots, N, \label{columy}
			\\  \text{colb}_i(\mathbf{X}) &= (\text{rowb}_i(\mathbf{X}^T))^T \in \mathbb{R}^{\sum_{j=0}^{k}p^j}, \,\,\,  i =1,2,\cdots, N.\label{X}
		\end{align}
		Employing matrix notation is possible to reduce the system of $k+1$ matrix equation to only one 
		\begin{align}\label{c11}
			\sum_{i=1}^{N} K_{h,i}(x) f(Y_i) \begin{bmatrix}
				I_n\otimes (X_i - x)^{\otimes 0}
				\\ I_n\otimes (X_i - x)^{\otimes 1}
				\\ \cdots
				\\ \cdots
				\\I_n\otimes (X_i - x)^{\otimes k}
			\end{bmatrix}^T = \sum_{i,j}^{N,k} \widehat{\beta}_j K_{h,i}(x) \left(I_n\otimes (X_i - x)\right)^{\otimes j}\begin{bmatrix}
			I_n\otimes (X_i - x)^{\otimes 0}
			\\ I_n\otimes (X_i - x)^{\otimes 1}
			\\ \cdots
			\\ \cdots
			\\I_n\otimes (X_i - x)^{\otimes k}
		\end{bmatrix}^T. 
	\end{align}
	To further simplify this expression, we can note that the Definition  of $X$ implies that 
	\begin{align}\label{row}
		\text{rowb}_i (I_n\boxtimes \mathbf{X}^T) &= \begin{bmatrix}
			I_n\otimes (X_i - x)^{\otimes 0}
			\\ I_n\otimes (X_i - x)^{\otimes 1}
			\\ \cdots
			\\ \cdots
			\\I_n\otimes (X_i - x)^{\otimes k}
		\end{bmatrix}^T,
		\\ \text{rowb}_j(\text{colb}_i(I_n\boxtimes \mathbf{X})) & = I_n\otimes (X_i-x)^{\otimes j} \label{rowcol},
	\end{align}
	therefore 
	\begin{align*}\label{c12}
		\sum_{i=1}^{N} K_{i,h}(x) f(Y_i) \text{rowb}_i (I_n\boxtimes \mathbf{X}^T) &=^{\text{by \ref{row} } } \sum_{i,j}^{N,k} \widehat{\beta}_j K_{i,h}(x) \left(I_n\otimes (X_i - x)^{\otimes j}\right)\text{rowb}_i (I_n\boxtimes \mathbf{X}^T)
		\\ \sum_{i=1}^{N} K_{h,i}(x) \text{colb}_i(f(\mathbf{Y})) \text{rowb}_i (I_n\boxtimes \mathbf{X}^T) &=^{\text{by \ref{columy}}} \sum_{i =1}^{N} K_{i,h}(x) \left(\sum_{j=0}^{k} \widehat{\beta}_j  \left(I_n\otimes (X_i - x)^{\otimes j}\right)\right)\text{rowb}_i (I_n\boxtimes \mathbf{X}^T)
		\\  f(\mathbf{Y}) (I_n\boxtimes \mathbf{W}) (I_n\boxtimes \mathbf{X}^T) &=^{\text{by \ref{algeidenblock}}} \sum_{i =1}^{N} K_{i,h}(x) \left(\sum_{j=0}^{k} \widehat{\beta}_j  \left(I_n\otimes (X_i - x)^{\otimes j}\right)\right)\text{rowb}_i (I_n\boxtimes \mathbf{X}^T)
		\\ f(\mathbf{Y}) (I_n\boxtimes \mathbf{W}) (I_n\boxtimes \mathbf{X}^T) &=^{\text{by \ref{rowcol}}} \sum_{i =1}^{N} K_{i,h}(x) \left(\sum_{j=0}^{k}  \text{colb}_j(\widehat{\beta}) \text{rowb}_j(\text{colb}_i(I_n\boxtimes \mathbf{X})) \right)\text{rowb}_i (I_n\boxtimes \mathbf{X}^T)
		\\f(\mathbf{Y}) (I_n\boxtimes \mathbf{W}) (I_n\boxtimes \mathbf{X}^T) &=^{\text{by  \ref{algeidenblock}}} \sum_{i =1}^{N} K_{i,h}(x) \widehat{\beta} \text{colb}_i(I_n\boxtimes \mathbf{X}) \text{rowb}_i (I_n\boxtimes \mathbf{X}^T)
		\\f(\mathbf{Y}) (I_n\boxtimes \mathbf{W}) (I_n\boxtimes \mathbf{X}^T) &= \widehat{\beta}\sum_{i =1}^{N} K_{i,h}(x) \text{colb}_i(I_n\boxtimes \mathbf{X}) \text{rowb}_i (I_n\boxtimes \mathbf{X}^T)
		\\f(\mathbf{Y}) (I_n\boxtimes \mathbf{W}) (I_n\boxtimes \mathbf{X}^T) &=^{\text{by  \ref{algeidenblock}}}  \widehat{\beta}(I_n\boxtimes \mathbf{X})(I_n\boxtimes \mathbf{W}) (I_n\boxtimes \mathbf{X}^T)
		\\f(\mathbf{Y}) (I_n\boxtimes (\mathbf{W} \mathbf{X}^T)) &= \widehat{\beta}(I_n\boxtimes (\mathbf{XWX}^T))
		\\\widehat{\beta}& = f(\mathbf{Y})(I_n\boxtimes (\mathbf{WX}^T(\mathbf{XWX}^T)^{-})).
	\end{align*}
	Now because $\widehat{\beta}_0$ is the first component of $\widehat{\beta} =(\widehat{\beta}_0,\cdots,\widehat{\beta}_k)$ we have to multiply $\widehat{\beta}$ by $I_n\boxtimes e_0$, therefore 
	\begin{align}
		\begin{split}
			\widehat{\beta}_0 &= \widehat{\beta} (I_n\boxtimes e_0)
			\\ & = f(\mathbf{Y})(I_n\boxtimes (\mathbf{WX}^T(\mathbf{XWX}^T)^{-})) (I_n\boxtimes e_0)
			\\ & = f(\mathbf{Y})(I_n\boxtimes (\mathbf{WX}^T(\mathbf{XWX}^T)^{-}e_0)) \\& = f(\mathbf{Y})( (\mathbf{WX}^T(\mathbf{XWX}^T)^{-}e_0)\otimes I_n). 
		\end{split}
	\end{align} 
\end{proof}
\section{Regularity Conditions}\label{regular}
\begin{enumerate}
	\item[R.1] The Kernel $K$ is compactly supported, bounded kernel such that $\int K_h(u)u^{\otimes 2} du = \mu_2(K)\operatorname{vec}(I_p)$, where $\mu_2(K)\ne 0$ is a scalar. In addition, all odd-order moments of $K$ vanish.  
	\item[R.2] The point $x$ is in $\operatorname{supp}(F_X)$. At $x$, $F_X$, is continuously differentiable and all second-order derivatives of $(fm)(X) = (f \circ m)(X)$ are continuous. Where $f$ is the isometry induced by the metric $g^f$ and $F_X$ is the density distribution of $X$. Also, $(fm)(x)>0$.
	\item[R.3] The sequence of bandwidth $h $ is such that $n^{-1}h\to 0$ as $n\to \infty$.
	
\end{enumerate}
\section{Proof of Thm. (\ref{cons})}\label{teoconsis}
\begin{proof}
	To demonstrate the consistency of $\widehat{\alpha}_0(x;1,h)$, we adopt a strategy that involves deriving the formula for the asymptotic bias of $\widehat{\beta}_0(x;1,h)$ and then using the Continuous Mapping Theorem \cite{van2000asymptotic} to establish the consistency of the former estimator. This is possible because both estimators are linked by the diffeomorphism $f$.  
	\subsection{Asymptotic bias of   $\widehat{\beta}_0(x;1,h)$}
	By Theorem (\ref{teorema3}) the estimator $\widehat{\beta}_0(x;1,h)$ have the following closed analytical expression 
	\begin{align}
		\widehat{\beta}_0(x;1,h) = f(\mathbf{Y})\left((\mathbf{WX}^T(\mathbf{XWX}^T)^-e_0)\otimes I_n\right),
	\end{align}
	then it follows that 
	\begin{align}\label{tay2}
		\mathbb{E}\left[\widehat{\beta}_0(x;1,h)|X_1,\cdots,X_N\right] = (fm)(X_1,\cdots,X_N)\left((\mathbf{WX}^T(\mathbf{XWX}^T)^-e_0)\otimes I_n\right),
	\end{align}
	where $(fm)(X_1,\cdots,X_N) = (f(m(X_1)),\cdots,f(m(X_N))\in\operatorname{M}_{n,nN}(\mathbb{R})$. Each component $(fm)(X_i)$  can be expanded using a second order Taylor expansion Eq. (\ref{taylorcoef})  therefore 
	\begin{align}
		(fm)(X_i) = (fm)(x) +\partial_x(fm)(X) (I_n\otimes (X_i-x)) +\frac{1}{2}\partial^2_x(fm)(X)(I_n\otimes (X_i-x)^{\otimes 2}) + r_i,
	\end{align}
	where $r_i$  is an error term. This Taylor expansion implies that we can use the following approximation for $(fm)(X_1,\cdots, X_N)$ 
	\begin{align}
		\begin{split}
			(fm)(X_1,\cdots,X_N) &= \begin{bmatrix}
				(fm)(X_1)
				\\(fm)(X_2)
				\\\cdots
				\\(fm)(X_N)
			\end{bmatrix}^T 
			\\&= \begin{bmatrix}
				(fm)(x) +\partial_x(fm)(X) (I_n\otimes (X_1-x)) +\frac{1}{2}\partial^2_x(fm)(X)(I_n\otimes (X_1-x)^{\otimes 2})
				\\(fm)(x) +\partial_x(fm)(X) (I_n\otimes (X_2-x)) +\frac{1}{2}\partial^2_x(fm)(X)(I_n\otimes (X_2-x)^{\otimes 2})
				\\\cdots
				\\(fm)(x) +\partial_x(fm)(X) (I_n\otimes (X_N-x)) +\frac{1}{2}\partial^2_x(fm)(X)(I_n\otimes (X_N-x)^{\otimes 2})
			\end{bmatrix}^T  +R
			\\& = \begin{bmatrix}
				(fm)(x) +\partial_x(fm)(X) (I_n\otimes (X_1-x))
				\\(fm)(x) +\partial_x(fm)(X) (I_n\otimes (X_2-x)) 
				\\\cdots
				\\(fm)(x) +\partial_x(fm)(X) (I_n\otimes (X_N-x)) 
			\end{bmatrix}^T +\begin{bmatrix}
			\frac{1}{2}\partial^2_x(fm)(X)(I_n\otimes (X_1-x)^{\otimes 2})
			\\\frac{1}{2}\partial^2_x(fm)(X)(I_n\otimes (X_2-x)^{\otimes 2})
			\\\cdots
			\\\frac{1}{2}\partial^2_x(fm)(X)(I_n\otimes (X_N-x)^{\otimes 2})\end{bmatrix} ^T +R,
	\end{split}
\end{align}
to continue, we have to take into account the following identities
\begin{align}
	\begin{split}
		\begin{bmatrix}
			(fm)(x) +\partial_x(fm)(X) (I_n\otimes (X_1-x))
			\\(fm)(x) +\partial_x(fm)(X) (I_n\otimes (X_2-x)) 
			\\\cdots
			\\(fm)(x) +\partial_x(fm)(X) (I_n\otimes (X_N-x)) 
		\end{bmatrix}^T & = \begin{bmatrix}
		(fm)(x) & \partial _x (fm)(X) 
	\end{bmatrix} \begin{bmatrix}
	I_n &  \cdots& I_n
	\\ I_n\otimes (X_1-x) & \cdots & I_n\otimes(X_N-x)
\end{bmatrix}
\\& = \begin{bmatrix}
	(fm)(x) & \partial _x (fm)(X) 
\end{bmatrix} (I_n\boxtimes \mathbf{X}),
\end{split}
\end{align}
and 
\begin{align}
	\begin{split}
		\begin{bmatrix}
			\frac{1}{2}\partial^2_x(fm)(X)(I_n\otimes (X_1-x)^{\otimes 2})
			\\\frac{1}{2}\partial^2_x(fm)(X)(I_n\otimes (X_2-x)^{\otimes 2})
			\\\cdots
			\\\frac{1}{2}\partial^2_x(fm)(X)(I_n\otimes (X_N-x)^{\otimes 2})\end{bmatrix} ^T & = \frac{1}{2} \partial^2_x(fm)(X)\begin{bmatrix}
			I_n \otimes (X_1-x)^{\otimes 2} , \cdots,  I_n \otimes (X_N-x)^{\otimes 2}
		\end{bmatrix}
		\\& =  \frac{1}{2} \partial^2_x(fm)(X)\left( I_n\boxtimes [(X_1-x)^{\otimes 2}, \cdots, (X_N-x)^{\otimes 2}]\right)
		\\& = \frac{1}{2} \partial^2_x(fm)(X)\left( I_n\boxtimes \Delta_x(X_1,\cdots,X_N)\right),
	\end{split}
\end{align}
where we denote $\Delta_x(X_1,\cdots,X_N) = [(X_1-x)^{\otimes 2}, \cdots, (X_N-x)^{\otimes 2}] \in \operatorname{M}_{p^2,N}(\mathbb{R})$.  Therefore from these identities, we conclude that 
\begin{eqnarray}\label{tay1}
	(fm)(X_1,\cdots,X_N)  \sim [(fm)(x)  \,\, \partial _x (fm)(X)] (I_n\boxtimes \mathbf{X}) + \frac{1}{2} \partial^2_x(fm)(X)\left( I_n\boxtimes \Delta_x(X_1,\cdots,X_N)\right).
\end{eqnarray}
From this point forward, we neglected the error term $R$ because multiplied at the right by  $\left((\mathbf{WX}^T(\mathbf{XWX}^T)^-e_0)\otimes I_n\right)$  and the resulting therm if of little order respect to the second term in the Taylor expansion of $(fm)(X_1,\cdots, X_N)$ in Eq. (\ref{tay1}).   To continue, we derived how the first term in Taylor expansion Eq. (\ref{tay1}) is affected. We multiply it by $\left((\mathbf{WX}^T(\mathbf{XWX}^T)^-e_0)\otimes I_n\right)$ 
\begin{align}
	\begin{split}
		&[(fm)(x)  \,\, \partial _x (fm)(X)] (I_n\boxtimes \mathbf{X})\left((\mathbf{WX}^T(\mathbf{XWX}^T)^-e_0)\otimes I_n\right) 
		\\&= [(fm)(x)  \,\, \partial _x (fm)(X)] (I_n\boxtimes \mathbf{X})\left(I_n\boxtimes(\mathbf{WX}^T(\mathbf{XWX}^T)^-e_0)\right)
		\\ &= [(fm)(x)  \,\, \partial _x (fm)(X)] \left(I_n\boxtimes(\mathbf{XWX}^T(\mathbf{XWX}^T)^-e_0)\right)
		\\ &= [(fm)(x)  \,\, \partial _x (fm)(X)] \left(I_n\boxtimes e_0\right)
		\\ & = (fm)(x).
	\end{split}
\end{align}
Therefore we conclude that 
\begin{align}\label{tay3}
	\mathbb{E}\left[\widehat{\beta}_0(x;1,h) - (fm)(x)|X_1,\cdots,X_N\right]& \sim  \frac{1}{2} \partial^2_x(fm)(X)\left(I_n\boxtimes(\Delta_x(X_1,\cdots,X_N)\mathbf{WX}^T(\mathbf{XWX}^T)^-e_0)\right).
\end{align}
Now we have to compute the expressions $N^{-1}\Delta_x(X_1,\cdots,X_N)\mathbf{WX}^T$ and $N^{-1}(\mathbf{XWX}^T)^-e_0$. Lets  start computing $N^{-1}\Delta_x(X_1,\cdots,X_N)\mathbf{WX}^T$
\begin{align}\label{tay6}
	\begin{split}
		N^{-1}\Delta_x(X_1,\cdots,X_N)\mathbf{WX}^T & = N^{-1}[(X_1-x)^{\otimes 2}, \cdots, (X_N-x)^{\otimes 2}] \operatorname{diag} (K_h(X_1-x),\cdots, K_h(X_N-x)) \begin{bmatrix}
			1 & (X_1 - x)^T
			\\ \cdots & \cdots
			\\1 & (X_N - x)^T
		\end{bmatrix}
		\\& = N^{-1}[(X_1-x)^{\otimes 2}, \cdots, (X_N-x)^{\otimes 2}]\begin{bmatrix}
			K_h(X_1-x) & K_h(X_1-x)(X_1 - x)^T
			\\ \cdots & \cdots
			\\ K_h(X_N-x) &  K_h(X_N-x)(X_N - x)^T
		\end{bmatrix}
		\\& = N^{-1}\begin{bmatrix}
			\sum_{i=1}^{N} K_h(X_i-x) (X_i-x)^{\otimes 2}, \sum_{i=1}^{N} K_h(X_i-x) (X_i-x)^{\otimes 2} (X_i-x)^T
		\end{bmatrix},
	\end{split}
\end{align}
and 
\begin{align}
	\begin{split}
		N^{-1}\sum_{i=1}^{N} K_h(X_i-x) (X_i-x)^{\otimes 2} (X_i-x)^T & = \int K_h(u) (hu)^{\otimes 2}(hu)^TF_X(x+hu)du + o_p(h^31_{p^2\times p})
		\\& = o_p(h^31_{p^2\times p}).
		\\N^{-1} \sum_{i=1}^{N} K_h(X_i-x) (X_i-x)^{\otimes 2} & = \int K_h(u) (hu)^{\otimes 2 }F_X(x+hu)du + o_p(h^21_{p^2\times 1})
		\\& = F_X(x) \int K_h(u) (hu)^{\otimes 2 } du + o_p(h^21_{p^2\times 1})
		\\& = F_X(x)h^2\mu_2(K) \operatorname{vec}(I_p) +o_p(h^21_{p^2\times 1}).
	\end{split}
\end{align}
where $1_{n,m}$ is a matrix on $\operatorname{M}_{n,m}(\mathbb{R})$ with all elements equal to one. Therefore  
\begin{align}
	N^{-1}\Delta_x(X_1,\cdots,X_N)\mathbf{WX}^T = [F_X(x)h^2\mu_2(K) \operatorname{vec}(I_p) +o_p(h^21_{p^2\times 1}), o_p(h^31_{p^2\times p})].
\end{align}
On the other hand, a well-known result from the literature is that \cite{ruppert1994multivariate}
\begin{align}
	(N^{-1} \mathbf{XWX}^T)^- = \begin{bmatrix}
		F_X(x)^{-1} +o_p(1) & J_F^T F_X(x)^{-2} +o_p(1_{1\times p })
		\\ - J_F F_X(x)^{-2} +o_p(1_{p\times1}) &  (h^2\mu_2(K)F_X(x)I_p)^{-1}+o_p(h^{-2}I_p)
	\end{bmatrix},
\end{align}
where $J_F$ is the gradient of the density  distribution $F(X)$. Therefore 
\begin{align}
	\begin{split}
		(\mathbf{XWX}^T)^-e_0 &= \frac{1}{N} \begin{bmatrix}
			F_X(x)^{-1} +o_p(1) & J_F^T F_X(x)^{-2} +o_p(1_{1\times p })
			\\ - J_F F_X(x)^{-2} +o_p(1_{p\times1}) &  (h\mu_2(K)F_X(x)I_p)^{-1}+o_p(h^{-1}I_p)
		\end{bmatrix}\begin{bmatrix}
		1\\O_{p\times 1}
	\end{bmatrix}
	\\& = \frac{1}{N} \begin{bmatrix}
		F_X(x)^{-1} +o_p(1)
		\\ - J_F F_X(x)^{-2} +o_p(1_{p\times1}) 
	\end{bmatrix}.
\end{split}
\end{align}
Therefore  
\begin{align}
	\begin{split}
		&\mathbb{E}\left[\widehat{\beta}_0(x;1,h) - (fm)(x)|X_1,\cdots,X_N\right] \sim  \frac{1}{2} \partial^2_x(fm)(X)\left(I_n\boxtimes(\Delta_x(X_1,\cdots,X_N)\mathbf{WX}^T(\mathbf{XWX}^T)^-e_0)\right)
		\\& = \frac{1}{2} \partial^2_x(fm)(X)\left(I_n\boxtimes\left(\begin{bmatrix}
			F_X(x)h^2\mu_2(K) \operatorname{vec}(I_p) +o_p(h^21_{p^2\times 1}), 
			& o_p(h^31_{p^2\times p})
		\end{bmatrix}\begin{bmatrix}
		F_X(x)^{-1} +o_p(1)
		\\ - J_F F_X(x)^{-2} +o_p(1_{p\times1}) 
	\end{bmatrix}\right)\right)
	\\& = \frac{1}{2} \partial^2_x(fm)(X)\left(I_n\boxtimes\left(h^2 \mu_2(K) \operatorname{vec}(I_p)+o_p(h^21_{p^2\times 1})\right)\right) 
	\\& = \frac{1}{2} \partial^2_x(fm)(X)\left(h^2 \mu_2(K) \operatorname{vec}(I_p)\otimes I_ n+o_p(h^21_{p^2\times 1}\otimes I_n)\right) 
	\\& = \frac{1}{2}h^2 \mu_2(K) \partial^2_x(fm)(X)\operatorname{vec}(I_p)\otimes I_n +o_p(h^2 I_n). 
\end{split}
\end{align}

From where we can conclude the statement of the Theorem  
\begin{align}\label{final}
	\mathbb{E}\left[\widehat{\beta}_0(x;1,h) - (fm)(x)|X_1,\cdots,X_N\right] =\frac{1}{2}h^2 \mu_2(K) \partial^2_x(fm)(X)\operatorname{vec}(I_p)\otimes I_n +o_p(h^2 I_n). 
\end{align}

\subsection{Consistency of the estimator $\widehat{\alpha}_0(x;1,h)$}
Upon taking the limit of Equation (\ref{final}), we can infer that the estimator $\widehat{\beta}_0(x;1,h)$ is consistent. Furthermore, as per Theorem (\ref{teorema3}), this estimator is linked to $\widehat{\alpha}_0(x;1,h)$ through the expression:

\begin{align}
	\widehat{\alpha}_0(x;1,h)=f^{-1}(\widehat{\beta}_0(x;1,h)),
\end{align}
where $f^{-1}$ is a diffeomorphism, owing to the fact that $f$ is both an isometry and a diffeomorphism. By virtue of the Continuous Mapping Theorem \cite{van2000asymptotic} and Theorem (\ref{teorema3}), along with the continuity of $f^{-1}$, we can establish the consistency of the estimator $\widehat{\alpha}_0(x;1,h)$.

\end{proof}
\end{appendix}
\bibliographystyle{unsrt}
\bibliography{sample.bib3}
\end{Backmatter}

\end{document}